\documentclass[%
 aip,
 amsmath,amssymb,
 reprint,%
]{revtex4-1}

\usepackage{graphicx}% Include figure files
\usepackage{dcolumn}% Align table columns on decimal point
\usepackage{bm}% bold math
\usepackage[utf8]{inputenc}
\usepackage[T1]{fontenc}
\usepackage{mathptmx}
\usepackage{etoolbox}
\usepackage{hyperref}
\makeatletter
\def\@email#1#2{%
 \endgroup
 \patchcmd{\titleblock@produce}
  {\frontmatter@RRAPformat}
  {\frontmatter@RRAPformat{\produce@RRAP{*#1\href{mailto:#2}{#2}}}\frontmatter@RRAPformat}
  {}{}
}%
\makeatother

\DeclareUnicodeCharacter{03B2}{$\beta$}
\DeclareUnicodeCharacter{2212}{$\text{-}$}
\DeclareUnicodeCharacter{03B4}{$\delta$}
\DeclareUnicodeCharacter{03B1}{$\alpha$}

\begin{document}

\preprint{AIP/123-QED}

\title{Utilizing {$(Al,Ga)_2O_3/Ga_2O_3$} superlattices to measure cation vacancy diffusion and vacancy-concentration-dependent diffusion of Al, Sn, and Fe in {$\beta\text{-}Ga_2O_3$}}
% Force line breaks with \\
\author{Nathan D. Rock}
\author{Haobo Yang}
\author{Brian Eisner}
\author{Aviva Levin}
\affiliation{Department of Materials Science and Engineering, University of Utah, Salt Lake City, UT 84112, USA}

\author{Arkka Bhattacharyya}
\author{Sriram Krishnamoorthy}
\affiliation{Department of Electrical and Computer Engineering, University of Utah, Salt Lake City, UT 84112, USA}
\affiliation{Materials Department, University of California, Santa Barbara, Santa Barbara, CA 93106, USA}

\author{Praneeth Ranga}
\affiliation{Department of Electrical and Computer Engineering, University of Utah, Salt Lake City, UT 84112, USA}

\author{Michael A Walker}
\affiliation{Department of Metallurgical and Materials Engineering, Colorado School of Mines,Golden CO, 80401, USA}

\author{Larry Wang}
\author{Ming Kit Cheng}
\author{Wei Zhao}
\affiliation{Eurofins EAG Material Science, Sunnyvale, CA, 94086, USA}

\author{Michael A. Scarpulla}
\affiliation{Department of Materials Science and Engineering, University of Utah, Salt Lake City, UT 84112, USA}
\affiliation{Department of Electrical and Computer Engineering, University of Utah, Salt Lake City, UT 84112, USA}
\email{mike.scarpulla@utah.edu}

\date{\today}% It is always \today, today,
             %  but any date may be explicitly specified

\begin{abstract}
Diffusion of native defects such as vacancies and their interactions with impurities are fundamental in semiconductor crystal growth, device processing, and long-term aging of equilibration and transient diffusion of vacancies are rarely investigated.  We used $(Al_xGa_{1-x)2}O_3/Ga_2O_3$ superlattices (SLs) to detect and analyze transient diffusion of cation vacancies during annealing in $O_2$ at $1000-1100 ^\circ C$.  Using a novel finite difference scheme for the diffusion equation with time- and space-varying diffusion constant, we extract diffusion constants for Al, Fe, and cation vacancies under the given conditions, including the vacancy concentration dependence for Al.  In the case of SLs grown on Sn-doped $\beta-Ga_2O_3$ (010) substrates, gradients observed in the extent of Al diffusion indicate that vacancies present in the substrate transiently diffuse through the SLs, interacting with Sn as it also diffuses.  In the case of SLs grown on (010) Fe-doped substrates, the Al diffusion is uniform through the SLs, indicating a depth-uniform concentration of vacancies.  We find no evidence in either case for the introduction of $V_{Ga}$ from the free surface at rates sufficient to affect Al diffusion down to ppm concentrations, which has important bearing on the validity of typically-made assumptions of vacancy equilibration.  Additionally, we show that unintentional impurities in Sn-doped $Ga_2O_3$ such as Fe, Ni, Mn, Cu, and Li also diffuse towards the surface and accumulate.  Many of these likely have fast interstitial diffusion modes capable of destabilizing devices over time, thus highlighting the importance of controlling unintentional impurities in $\beta-Ga_2O_3$ wafers.
\end{abstract}

\maketitle

\section{\label{Intro}Introduction}

$\beta{\text-}Ga_2O_3$ is  an ultra-wide band gap (UWBG) semiconductor with potential for superior performance in power electronics as well as in solar blind UV photodetectors and in transparent contacts for photovoltaics.  \cite{pearton_review_2018, green_-gallium_2022,tsao_ultrawide-bandgap_2018}.    To further advance this technology, a thorough understanding of the defects present and their transport is needed including during single crystal growth, epitaxial thin film growth, and during other fabrication steps.

Many groups have observed an increase in resistivity of n-type $\beta{\text-}Ga_2O_3$ after $O_2$ annealing\cite{oshima_vertical_2008}.  This formation of an insulating surface layer allows facile formation of UV photodetectors. This phenomenon was originally attributed to the elimination of $V_O$ which were presumed to act as shallow donors, before refined DFT calculations indicated that $V_O$ was a deep donor and thus could not contribute directly to n-type conductivity\cite{varley_oxygen_2010}.  Oshima, et al. documented the $\sqrt{t}$ kinetics of the insulating layer thickness formed during $O_2$ annealing and demonstrated that an $SiO_2$ cap could prevent it\cite{oshima_formation_2013}.  This indicates that direct exposure of the $Ga_2O_3$ surface to oxygen does control the compensation of n-type conductivity in some way.  Meanwhile positron annihilation experiments have demonstrated the presence of $V_{Ga}$ in both bulk crystals and thin films\cite{tuomisto_defect_2013,korhonen_electrical_2015}, and that increases in their numbers and or charge states accompany $O_2$ annealing   , which makes them a strong candidate for n-type compensation\cite{ingebrigtsen_impact_2019, jesenovec_gallium_2021}.  Many mechanisms, involving oxygen, $V_{Ga}$, $V_O$, and n-type dopants (Sn and Si) have been proposed, including those which involve defect complexes instead of bare $V_{Ga}$ and $V_O$\cite{frodason_multistability_2021,karjalainen_interplay_2021,kyrtsos_migration_2017,sun_oxygen_2021}.  Studying the diffusion of $V_{Ga}$ should elucidate what mechanisms were at play, however vacancies are  much more difficult to observe directly than e.g. impurity atoms.

With the exception of the direct interstitial impurity mechanism, self- and tracer-diffusion is mediated by native defects such as vacancies or interstitials.\cite{de_souza_oxygen_1998,topfer_point_1995,howard_random-walk_1966,klugkist_tracer_1995,mehrer_diffusion_2007,tsao_ultrawide-bandgap_2018}.    Modern computation and experiment have established Ga vacancies and their complexes as the most plentiful native defects in n-type $\beta{\text-}Ga_2O_3$. Substitutional isovalent alloying cations in oxides having ionic radii not significantly less than the host atom – e.g. Al in $Ga_2O_3$ – should exhibit primarily vacancy-mediated mechanisms under oxygen-rich conditions favoring cation vacancies and suppressing cation interstitials.  Interstitial mediated mechanisms may be possible under metal-rich conditions favoring cation interstitials.  The local diffusion constant of isotopic or chemical tracers is proportional to the local concentration of mediating native defects.  Thus for vacancy-hopping the tracer diffusion “constant” can vary in space and time with the concentration of mediating vacancies and different crystals may exhibit different diffusion constants by virtue of their vacancy content.  This in turn is affected by dopant impurities through Fermi level effects and their thermal and chemical history.   For dopant ions that control the local Fermi level, local equilibrium of native defects can result in that dopant’s diffusion constant being dependent on its own concentration   \cite{johansen_aluminum_2015,frodason_diffusion_2023, bracht_diffusion_2000,gosele_point_1991}.  The most general microscopic formulation of tracer diffusion problems is coupled reaction-diffusion of defects and complexes wherein the different tracer-native defect complexes have different diffusion coefficients than the isolated species\cite{hobbs_chapter_2015, krasikov_beyond_2018}.    The fraction of time, or population fraction, of complexes relates to the binding energy and connects to the correlation coefficient of random walk theory  \cite{mehrer_diffusion_2007}. Interestingly, our assumption that the Al impurity diffusion constant in Fick’s laws is proportional to vacancy concentration is not on its face identical to introducing non-zero but constant off-diagonal coefficients within Onsager’s transport formalism\cite{onsager_reciprocal_1931}.  

Herein, isothermally anneal superlattices of isovalent Al tracer concentration spikes in $\beta$-Ga2O3 to investigate the interdiffusion of Al, diffusion of cation vacancies, and diffusion of Fe or Sn impurities.  Commonly it is assumed (and in careful experiments, ensured) that diffusion-mediating native defects equilibriate with local impurity concentrations rapidly compared to the experimental timescale.  Rapid equilibration can be assumed in samples containing dense sources and sinks, however in high-perfection crystals like those in this study, equilibration may be much slower and limited by injection at or transport from remote surfaces.  Experiments in which native defect processes occur on similar timescales to tracer or impurity diffusion translate to time- and space-varying vacancy concentrations and thus impurity diffusion constants.  

Herein, Al diffusion in superlattices (SLs) epitaxially grown on Fe substrates occurs with at least spatially-uniform vacancy density, while in samples grown on Sn-doped wafers we show strong evidence for transient diffusion of cation vacancies out of the substrate.  Using a novel modification of the Crank-Nicholsen finite differences scheme, we infer the diffusion of the vacancies mediating Al diffusion.  The final Al profiles in this work record a weighted time average of $[V_{Ga}]$ at each depth and the $D_{V_{Ga}}$ we extract should also be understood as a time-averaged value in the presence of certain Sn (donor) or Fe (acceptor) concentrations and small amounts of Al.  Our experiments thus set lower bounds on the true $D_{V_{Ga}}$ and Ga self-diffusion constant for single crystalline $\beta\text{-}Ga_2O_3$.  We determine the diffusion constants for Al including its dependence on local $V_{Ga}$ concentration, for $V_{Ga}$ (likely in the form of $Sn_{Ga}-V_{Ga}$ complexes), and for Fe under conditions of low vacancy concentration and low electron density.  Finally, we surreptitiously discovered that many other cation species such as Fe, Mn, Ni, Li, and Cu also diffused from the Sn-doped substrates and accumulated near the free surface.

\section{\label{sec:methods}Experimental and Numerical Methods} 
Superlattices (SLs) consisting of 6 or 9 alternating layers of 200 nm unintentionally doped (UID) $\beta{\text-}Ga_2O_3$ and 10-15 nm $Al_{2x}Ga_{2(1-x)}O_3$ (AlGO) where $x \cong 0.05$. were epitaxially grown using organometallic vapor phase epitaxy (OMVPE) in an Agnitron Agilis 100 system on (010) Sn or Fe doped $\beta{\text-}Ga_2O_3$ substrates obtained from Novel Crystal Technology (NCT).  The vendor reports Fe or Sn concentrations in the mid $10^{18} cm^{-3}$, and etch pits from nanovoids and dislocations are $<10^5 /cm^2$.  The lowest practical Al concentration of $\approx5$ at.\% was used and layers below the critical thickness were used in order to prevent introduction of misfit dislocations.  SLs always began and ended with 200 nm $Ga_2O_3$ spacers.  UID layers grown under the same conditions exhibit n-type doping in the low $10^{16} /cm^3$ range.  Annealing was carried out for times between 2 and 80 hours with samples face up in a quartz tube furnace in 1 atm flowing $O_2$ after acid cleaning the tube and the quartz or sapphire plates on which samples sat.  During annealing, temperature was ramped at $10 ^\circ C/min$ and then held constant for the indicated time, after which, it was allowed to freely return to room temperature.  

Dynamic secondary ion mass spectroscopy (d-SIMS) profiles were collected before and after annealing on each individual sample by EAG Eurofins.  Depth was calibrated by measured crater depths and concentrations using ion implanted composition standards.  Comparing multiple SIMS profiles from each sample and between samples as well as x-ray diffraction Pendellosung fringes  allowed us to quantify the uncertainty in depth scales, resulting primarily from growth rate variation, of at most $\Delta x=\pm 6\%$.   This translates to relative uncertainty in extracted diffusion constants of order $(\Delta x)^2$=0.4$\%$, which we expect is much smaller than uncertainties from sample to sample variations in doping and defects.  The annealing temperature accuracy at the location of the samples was measured to be within a few degrees of nominal.    

Isovalent aluminum and gallium have similar ionic radii, therefore we assume a substitutional, vacancy mediated diffusion mechanism for Al.   Somewhat surprisingly, numerical treatment of the case of non-steady-state vacancy-mediated diffusion, which amounts to assuming that the tracer diffusion “constant” varies in space and time, is not widely considered.  Thus, a novel numerical approach based on the Crank-Nicholson scheme ($2^nd$ order centered finite differences in space, forward and linear in time) was developed.  The local Al diffusion constant $D_{Al}$ is assumed to be Arrhenius activated and proportional to the local concentration of cation vacancies (shorthanded as $[V_{Ga}]$ in $\#/cm^3$ acknowledging the small Al content) and thus takes the form 

\begin{eqnarray} \label{eq:Doo}
    D_{Al}&= D_{o}\cdot exp\left(\frac{-E_a}{k_B T}\right) =D^*\cdot [V_{Ga}] \cdot exp\left(\frac{-E_a}{k_B T}\right) \nonumber \\
    &\qquad= D_{oo}(T) \cdot [V_{Ga}]	
\end{eqnarray}

in which $D_o (cm^2/s)$ is the conventional prefactor, $D^* (cm^5/s)$ is a constant for all temperatures, $E_a$ is the Al hopping activation energy given a neighboring vacancy, $k_B$ is Boltzmann’s constant, T is absolute temperature, and $D_{oo}(T) = D^* exp\left(\frac{-E_a}{k_B T}\right)$ is a constant at each temperature.  Fick’s $2^nd$ law for isothermal tracer Al diffusion becomes

\begin{eqnarray} \label{eq:derrivation}
    \frac{\partial[Al]}{\partial t}&=\frac{\partial}{\partial x} \left(D_{Al}  \frac{\partial[Al]}{\partial x}\right)=\frac{\partial D_{Al}}{\partial x}  \frac{\partial[Al]}{\partial x}+D_{Al}  \frac{\partial^2 [Al]}{\partial x^2} \nonumber \\ 
    &\qquad=D_{oo} \biggl\{\frac{\partial[V_{Ga}]}{\partial x}  \frac{\partial[Al]}{\partial x}+[V_{Ga}]  \frac{\partial^2 [Al]}{\partial x^2} \biggr\}
\end{eqnarray} 
Not assuming spatial uniformity of the diffusion constant introduces the cross-derivative term involving gradients of both [Al] and $[V_{Ga}]$.  The $2^{nd}$ order space central difference, first order forward time discretization of this equation on a uniform grid of spacing $\Delta x$  is

\begin{eqnarray} \label{eq:finitedifference}
     \frac{u_N^{j+1}-u_N^j}{\Delta t}&=\frac{1}{8\Delta x^2 }  [ u_{N-1}^j (D_{N-1}^j+4D_N^j-D_{N+1}^j )\nonumber \\ \nonumber
     &\qquad+u_{N-1}^{j+1} (D_{N-1}^{j+1}+4D_N^{j+1}-D_{N+1}^{j+1} )\\ \nonumber
    &\qquad-8 (u_N^j D_N^j+u_N^{j+1} D_N^{j+1} )\\ \nonumber
    &\qquad+u_{N+1}^j (-D_{N-1}^j+4D_N^j+D_{N+1}^j )\\
    &\qquad+u_{N+1}^{j+1} (-D_{N-1}^{j+1}+4D_N^{j+1}+D_{N+1}^{j+1} ) ]
\end{eqnarray}

in which the space and time grids are indexed respectively by the subscript (j) and superscript (N) and $\Delta x$ and $\Delta t$ are the x and t intervals.  Dirichlet and von-Neuman boundary conditions were applied for Al at the deepest points in the substrate and surface respectively.  To simplify simulation, which we implemented in Matlab\cite{noauthor_matlab_2022}, we used analytical solutions for diffusion of $[V_{Ga}]$ which is tantamount to assuming vacancy diffusion does not depend on Al concentration.  This assumption was justified by the chemical similarity of Al and Ga and the small maximum Al concentration. We find the best balance of assumptions and distinguishable differences in modeled profiles by assuming depth-constant (but distinct) initial $[V_{Ga}]$ conditions in the OMVPE grown layers and the substrates.  Because of the high dynamic range of the SIMS data with low uncertainty (meaning that both the peaks and valleys of the Al profiles are measured with good signal to noise ratios), we chose to use the square difference of $log_{10}([Al])$ between measured and simulated profiles (summed over the J points in each profile) as our goodness of fit (GOF) parameter:  

\begin{equation} \label{eq:GOF}
GOF = \frac{1}{J-3}\sum_{j=0}^J\biggl( log_{10}\bigl[Al_{model}(j)\bigr]-log_{10}\bigl[Al_{SIMS}(j)\bigr] \biggr)^2
\end{equation}

$D_{V_{Ga}}$, the product$ [V_{Ga}]_{subs}\cdot Doo$ and the ratio $[V_{Ga}]_{epilayer}/[V_{Ga}]_{subs}$ are the 3 free parameters (degrees of freedom) of each scenario directly-extractable from the simulation/fitting and robustly determined at each temperature.  Rather narrow ranges of parameter values reproduce the Al profiles especially for the samples grown on Sn substrates.  If we assume a value for $[V_{Ga}]$ in the substrates (discussed below), minimizing the GOF across many diffusion simulations allows us to extract $[V_{Ga}]$  in the epilayer, $D_{oo}$ for the Al at each $T_{anneal}$, and $D_{V{Ga}}$ at each T.  Any difference between actual and assumed $[V_{Ga}]_{subs}$ will affect $[V_{Ga}]_{epilayer}$ and $D_{oo}$; if the actual concentration is later found to be 10x higher than assumed in our analysis, then Doo would actually be 10x lower but $[V_{Ga}]_{epilayer}$ also 10x higher..  Uncertainties in the final extracted parameter values were estimated from the curvature of the GOF with respect to each parameter.  Further discussion of numerical methods and error propagation are in the Supplemental Materials.    

\section{Expectations of Vacancy Initial Conditions and Initial Observations on Al Diffusion}

First we discuss expectations for the initial conditions of concentrations of $V_{Ga}$ in the substrates and superlattices and then present and initially assess the SIMS data.  Intuition for a completely ionic crystal suggests $V_{Ga}$ should exist only in the $3^-$ charge state and thus have concentration scaling $\propto(\frac{n}{n_i})^3$ or $\propto exp\left(\frac{-3E_f}{k_BT}\right)$\cite{jacobs_f_1974}.  Modern density functional theory (DFT) with hybrid functionals accurately predicts that defect chargestates may change if $E_f$ moves above or below charge transition levels and can account for complexation with other defects\cite{varley_oxygen_2010,varley_first-principles_2020}.  At equilibrium with all other factors held constant, the concentration of $V_{Ga}$ should be proportional to the n-type doping concentration, however differences in chemical potentials (e.g. $p_{O_2}$) and temperatures during bulk crystal and OMVPE growth will modify this ratio.    

A number of experiments have\cite{von_bardeleben_high_2023,jesenovec_gallium_2021,tuomisto_defect_2013,tuomisto_identification_2021,karjalainen_interplay_2021,korhonen_electrical_2015} detected and studied cation vacancies, however the most credible \textit{quantification} of $V_{Ga}$ or its complexes in Sn-doped wafers comes from the $E_c$-2.0 eV signal measured by deep level optical spectrscopy (DLOS).  DLOS measurements on different but very similar OMVPE layers report a concentration of $\approx 10^{15} /cm^3$ for the $E_c$-2.0 eV defects\cite{ghadi_full_2020}, while DLOS measurements of $E_c-2.0$ eV defects in NCT Sn-doped bulk crystals are reported at $0.8-1.2x10^{16}   /cm^3$ for wafers matching  the nominal concentration of those used herein, $[Sn]=3.5-5x10^{18} /cm^3$\cite{johnson_unusual_2019}.  For analysis we assume $[V_{Ga}]=10^{16} /cm^3$ vacancies in the Sn-doped substrates.

\begin{figure*}
    \centering
    \includegraphics[width=\textwidth]{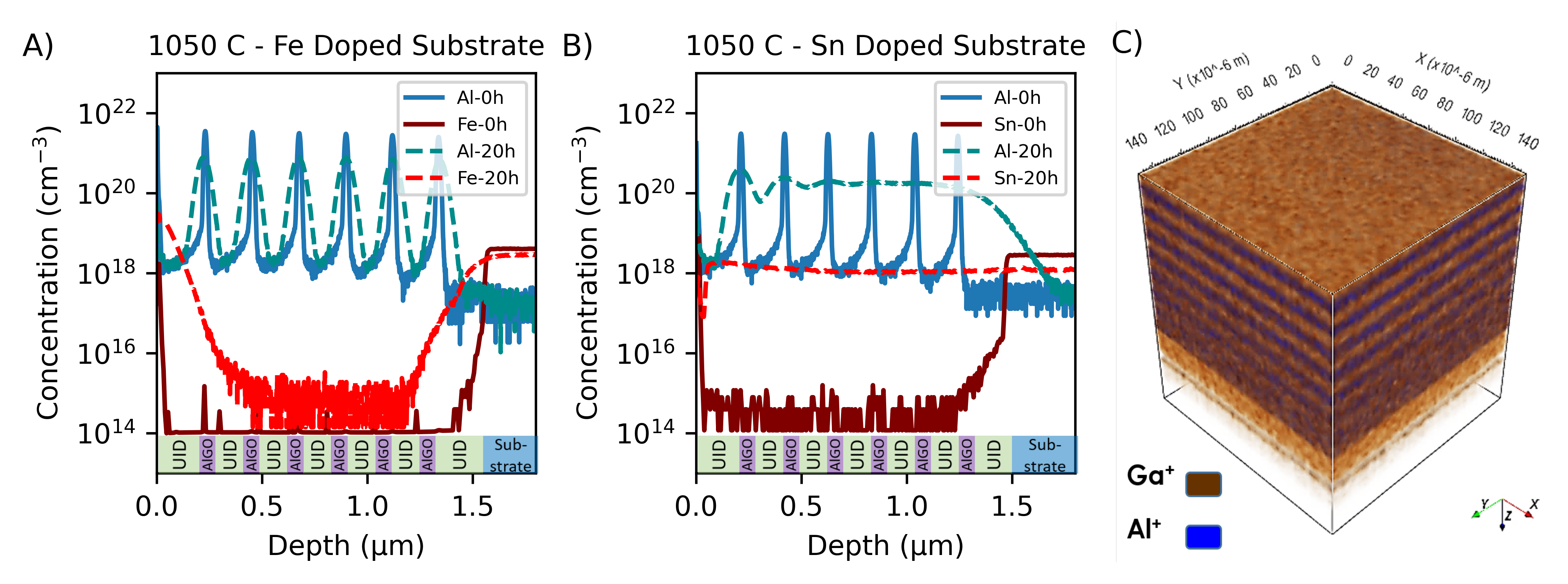}
    \caption{Pre- and post- annealing quantified Al, Sn, and Fe SIMS profiles for superlattices grown on (a) Sn-doped and (b) Fe-doped substrates and annealed 20 hrs in 1 atm flowing O2 at $1050^\circ C$. (c) 3D render of ToF SIMS of SL after annealing at $1100^\circ C$ for 20 hours (see Fig \ref{fig:isochronal})}
    \label{fig:SnvFeSchematic}
\end{figure*}

 While it is possible that DLOS may not detect some vacancies bound in complexes, the near-unity doping efficiency measured for Sn in bulk substrates in the $10^{18} /cm^3$ range indicates that $[V_{Ga}]$ can not be much larger.  Unfortunately, junction capacitance based defect spectroscopies like DLTS or DLOS can not be performed on insulating samples e.g. Fe-doped wafers or oxygen annealed ones.  All substrates were grown at $T_{melt} (k_BT\approx 0.18 eV)$ by EFG in reduced $p_{O_2}$ conditions\cite{galazka_czochralski_2010, kuramata_high-quality_2016}.  In Fe-doped $Ga_2O_3$, the Fermi level is pinned at the $Fe^{3^{+}/2^{+}}$ charge transition level corresponding to the 0/- transition of the $Fe_{Ga}$ acceptor located very near $E_c$-0.8 eV by DLTS and related techniques\cite{zhang_deep_2016, ingebrigtsen_iron_2018}.  Sn-doped substrates are degenerately doped or very close to it, thus $E_f\approx E_c$.  For both of these $E_f$ values, isolated $V_{Ga}$ should be in the $q=3^-$ charge state; thus $[V_{Ga}^{3-}]$ in the Sn-doped substrates should be higher than in the Fe-doped by a factor of $exp\left(\frac{3 \times 0.8 eV}{0.18 eV}\right)\approx 10^6$ if $[V_{Ga}]$ in both substrate types equilibriate at $T_{1}$=$T_{melt}$.  If $V_{Ga}$ stay equilibrated to a $T_1$ temperature lower than the freezing temperature, this ratio would increase, while if most of the $V_{Ga}$ are in $(Sn_{Ga}-V_{Ga})^{2-}$ complexes or $(V_{Ga}-V_O)^-$ divacancy complexes, this ratio would be approximately $10^4$ or $100$ respectively.  While the unintentional doping in our OMVPE layers grown at $800 ^\circ C$ is in the low $10^{16}$, it is very difficult to predict the ratio of $[V_{Ga}]$ in the SLs to those in the substrates because of the very different growth conditions.  Also, in separate experiments we have carried out, we find that annealing Sn-doped wafers in 1 atm $O_2$ at $1050^\circ C$ requires approximately two weeks for full equilibration; thus $[V_{Ga}]$ will not equilibrate in the full wafer during OMVPE growth.  Thus, we consider ratio of $[V_{Ga}]$ in the SL to that in the substrate to be an unknown parameter to be determined in the fitting.     Note that the substrate may not equilibrate its $[V_{Ga}]$ during OMVPE growth, so the ratio between substrate and SL may not have this value.  Despite the nominally-identical growth of the superlattices, the initial conditions for the $V_{Ga}$ or complexes mediating Al diffusion may be different from run to run, for those grown on Fe or Sn doped substrates, and from individual substrate to substrate. 

Figure \ref{fig:SnvFeSchematic}a,b shows Al and dopant SIMS profiles before and after annealing for 20 hours in 1 atm $O_2$ at $1050 ^\circ C$ for two identical superlattices grown on Sn- or Fe-doped substrates.  For the SLs on Fe-doped substrates, Al diffuses without any detectable gradient in the extent of diffusion of the $AlGa_2O_3$ layers in the superlattices.  Meanwhile, for SLs on Sn-doped substrates the extent of Al diffusion is greater for AlGaO layers closer to the substrate than the surface, and overall the Al diffusion is much faster.  This is consistent with a larger $[V_Ga]$ in the Sn-doped substrates compared to the SLs and to that in the Fe-doped substrates which diffuses through the SLs towards the surface on a non-negligible timescale compared to the annealing time.  For both types of samples, we find no evidence for vacancy creation or in-diffusion from the free surface at concentrations sufficient to affect Al diffusion from about 5 at.\% down to $10^{18} /cm^3$.  Such an effect could be taking place, but apparently at concentrations significantly lower than the estimated $10^{16} /cm^3$ from the Sn-doped substrates.  In the case of Fe-doped substrates, $V_{Ga}$ diffusion could be much faster than the time required for Al diffusion at the given $V_{Ga}$ concentration such that it does equilibrate quickly compared to the experimental time.  As discussed later, $Sn_{Ga}$ donors form complexes with acceptor-like $V_{Ga}$ but acceptor-like $Fe_{Ga}$ do not; therefore we should expect isolated $V_{Ga}$ to diffuse faster through SLs grown on Fe-doped substrates. 

\section{\label{sec:Fe}Analysis of Al and Fe Diffusion for SLs on Fe-Doped Substrates}

For SLs on Fe-doped substrates, uniform $[V_{Ga}]$ in the SL and substrate eliminates the cross-derivative term in Eq. \ref{eq:derrivation} for the Al diffusion, reducing it to the usual form of Fick’s 2nd law.  We tried many scenarios of different [VGa] in SL and substrate but the best agreement with experiments was found for the case of the same, low concentrations.  Evolving the initial to final Al SIMS profiles for the sample annealed at $1050^\circ C$ allows us to extract a value of  $D_{Al}=D_{oo}(T)\cdot [VGa]=3.5x10^{-18} cm^2/s$, assuming $[V_{Ga}]=10^{14} /cm^3$, i.e., $\approx 100$ times less than the substrate value used for Sn doped substrates. 

The diffusion of Fe itself appears to be complex, probably involving at least two modes such as both interstitial and substitutional.  The difference between the integrated Fe concentration in the epi-layer, before and after annealing, is $\approx 8.4x10^{13} /cm^2$.  Of this, $8.0x10^{13} /cm^2$ accumulated near the surface, which we have shown to be neither a SIMS artifact or a result of vapor transport during annealing (See supplemental for details).  This effect was reproduced via dynamic and Time of Flight (TOF) SIMS, across multiple samples, and is also reproduced on Sn doped substrates where [Fe] is initially considerably lower in the substrate.  The remaining $4x10^{12} /cm^2$ Fe that diffused only short distances into the SL region apparently diffuses via a second mode.    Assuming this corresponds to diffusion of Fe into a long half-space without complexation yields the analytical solution to Fick’s $2^{nd}$ law\cite{crank_mathematics_1956}:

\begin{equation} \label{eq:erfsolution-Fe}
    C(x,t)=\frac{C_{subs}-C_{SL}}{2} \left(1+erf\left[ \frac{x-x_L}{\sqrt{Dt}} \right]\right)
\end{equation}

in which $C_{subs},C_{SL}$ are the initial [Fe] concentrations in the substrate and SL respectively, $x_L$ is the depth of the SL/substrate interface, and $D$ is the diffusion constant for free Fe.  The best fit of this equation, excluding the [Fe] near the surface, yields a diffusion constant of $D_{Fe}=6x10^{-16} cm^2/s$.  The diffusion of Fe in $Ga_2O_3$ certainly warrants additional experimental and theoretical scrutiny, including Fermi level effects.  Fe is present at appreciable concentrations in most if not all $Ga_2O_3$ bulk crystals as an impurity introduced from raw materials or Ir crucibles and dies \cite{mccloy_growth_2022}.  

\section{Diffusion of Al in AlGa2O3/Ga2O3 SLs on Sn-Doped Substrates} \label{sec:AlinSn}

\begin{figure*}
    \centering
    \includegraphics[width=\textwidth]{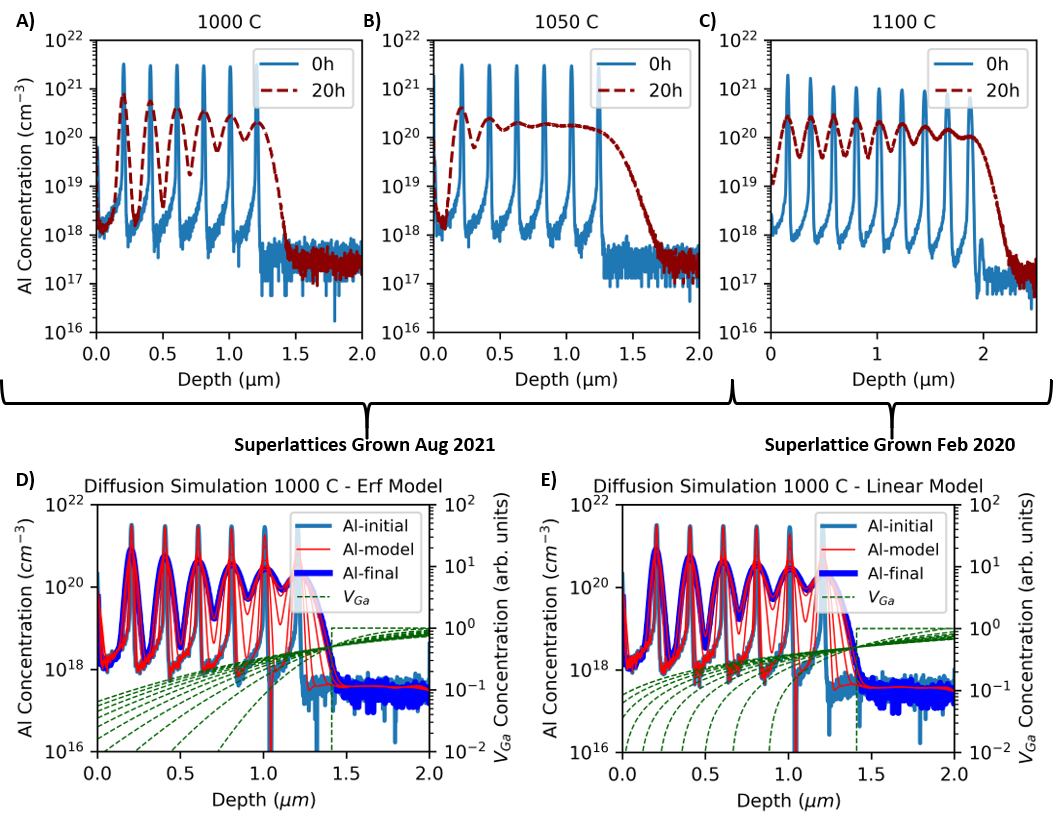}
    \caption{(A-C) SIMS profiles of Al before (blue) and after (red) annealing for 20 hours at the indicated temperatures.  (D-E) initial and final SIMS data from $1000^\circ C$ sample with results of finite differences simulation shown in red for 2 hour increments under the assumption that $V_{Ga}$ diffuses as $erf(x,t)$ (D) or as piece-wise linear solution to the $Sn_{Ga}-V_{Ga}$ complex reaction-diffusion (E)  A and B were grown in the same OMVPE growth, while (C) was grown in a different run.  The initial SIMS Al  profiles were used as initial conditions for the finite differences code with space- and time-varying diffusion constants and evolved to the final time; computed profiles are shown in red for increments of 2 hours.  Goodness of fit values were computed for the final profiles sweeping a large range of the model parameter space in order to determine the best values of $V_{Ga}$ initial step-function profiles, $V_{Ga}$ diffusion constant, and proportionality constant between $[V_{Ga}]$ and $D_{Al}$.  The assumed initial $V_{Ga}$ step function profile is evolved using either the $erf()$ assumption or the piece-wise-linear approximation to the reaction diffusion of complexes, as shown in dashed green.  The apparent decrease in the extent of Al diffusion between B) grown at $1050^\circ C$ and C) grown at $1100^\circ C$ is accounted for by a difference in the initial $[V_{Ga}]$ in the substrate, as explained in Sec. \ref{sec:AlinSn}}
    \label{fig:isochronal}
\end{figure*}

Figure \ref{fig:isochronal} a-c) present initial and final SIMS profiles for [Al] for samples annealed at $1000$, $1050$, and $1100^\circ C$ for 20 hours.  The extent of Al diffusion within each SL is greater near the substrate than near the surface indicating that the local concentration of $[V_{Ga}$] was non-uniform and greater near the substrate for a significant portion of the annealing time.  Samples A and B were grown in the same OMVPE growth run, while C was grown in a different run; this is important for understanding the apparently lesser degree of Al diffusion despite this sample’s higher annealing temperature, as discussed in more detail below.  Figure \ref{fig:isochronal} d and e reproduce the data from a) but also overlay simulated Al concentration profiles at 2 hour intervals (red solid) for the best-fit parameters within two assumptions for $V_{Ga}$ diffusion (green dashed) described in the following.

Diffusion parameter estimation from each experiment was accomplished by evolving the measured initial Al SIMS profile forward in time under many scenarios and fitting the final SIMS profiles.  Each tested scenario is defined by Eqs. \ref{eq:Doo}-\ref{eq:derrivation}, assumed initial $[V_{Ga}]$ in SL and substrate, $[V_{Ga}]$ diffusion behavior explained below, and values for $D_{oo}$ for Al and $D_{V_{Ga}}$.  It is clear for all Sn-doped samples that the initial $[V_{Ga}]$ in the SLs must be at least 10 times lower than in the substrates which agrees with intuition given the higher substrate n-type doping.  The effects of $D_{V_{Ga}}$ and the initial [$V_{Ga}$] ratio on the broadening and peak-to-valley [Al] ratio for each Al-containing layer in the SLs are sufficiently distinct that we can determine their values with low uncertainty within each scenario.  
As shown in Figure \ref{fig:isochronal} d), we first tested the hypotheses that a) $V_{Ga}$ are conserved and b) that they diffuse freely without interactions from the Sn-doped substrate through the SLs according to Eq. \ref{eq:erfsolution-Fe}\cite{crank_mathematics_1956}: 

%\begin{flalign} \label{eq:Triangle}
 %   &{Sn}_{Ga}V_{Ga}\left(x,t\right)= &&
%\end{flalign}
\begin{equation} \label{eq:Triangle}
    C\left(x,t\right)=
    \begin{cases}
            C_{subs} &\text{if } C\left(x,t\right)>C_{subs} \\
            C_{epi} &\text{if } C\left(x,t\right)<C_{epi} \\ 
            \frac{C_{max}}{2}\left(1+\frac{x-x_L}{\sqrt{Dt}}\right) &\text{if }C_{epi}\geq C\left(x,t\right).. \\
            &\text{and }C\left(x,t\right)\geq\ C_{subs}  
    \end{cases}
\end{equation}

in which $C_{subs},C_{SL}$ are the initial vacancy concentrations in the substrate and SL respectively, $x_L$ is the depth of the SL/substrate interface, $D$ is the diffusion constant for free vacancies, and t is the isothermal annealing time.  In the supplemental materials we explore the issue of $V_{Ga}$ conservation and implications for assumptions that vacancies equilibrate in other single crystal diffusion experiments.  Our simulations reproduced the Al SIMS data fairly well indicating that these two assumptions capture the gross features of $V_{Ga}$ diffusion in our experiments; no generation of $V_{Ga}$ in the SLs or introduction from the free surface are indicated at levels detectable in these experiments.

However, theoretical and experimental evidence\cite{frodason_diffusion_2023,varley_first-principles_2020} suggest that $V_{Ga}$ in Sn doped substrates dynamically complex with a small fraction of the available Sn ([$V_{Ga}$]<<[$Sn_{Ga}$]).  Complexed $V_{Ga}$ diffuse slower than free $V_{Ga}$, while $Sn_{Ga}$ diffuse only while complexed or correlated with a vacancy \cite{mehrer_diffusion_2007} .  Critically for assessing how this will affect Al diffusion, the $Sn_{Ga}-V_{Ga}$ complexation reaction’s equilibrium rate constant was determined to be small.  [Sn] is roughly 100-1000 times higher than $[V_{Ga}]$ in the substrate, which ensures that the majority of $V_{Ga}$ is on average bound in $Sn_{Ga}-V_{Ga}$ complexes.  However, the small equilibrium rate constant ensures that individual $Sn_{Ga}-V_{Ga}$ complexes are constantly dissociating and reforming.  Therefore, $V_{Ga}$ in the vicinity of Al has a high likelihood of assisting in Al diffusion even when the $V_{Ga}$ is diffusing as a complex.  Because any ionized donor will have Coulomb attraction to cation vacancies, the effective diffusion constant for $V_{Ga}$ and thus for cation self-diffusion in the presence of donors such as Sn should be smaller the value in intrinsic or acceptor-doped $\beta\text{-}Ga_2O_3$ (e.g. Fe-doped).  The degree of retardation of cation vacancy diffusion will depend on the concentrations of donors and vacancies and thus be variable from experiment to experiment and vs depth and time within any diffusion experiment.

Figure \ref{fig:isochronal} e) shows fitting of one data set using a modified assumption for the evolution of $[V_{Ga}](x,t)$, taking into account its complexation with $Sn_{Ga}$.  Frodason et al.\cite{frodason_diffusion_2023} studied the diffusion of Sn from (001) Sn-doped substrates into UID HVPE-grown $Ga_2O_3$ for temperatures $1050$ to $1250 ^\circ C$ and fit them with a diffusion-reaction model.  A key assumption of their work was that $[V_{Ga}]$ equilibrated quickly by unspecified generation or transport mechanisms with the local Fermi level set by the local doping – i.e. that $\int[V_{Ga}]dx$ was not conserved.  

\begin{table*}
\caption{\label{tab:resultssummary}Effective Al diffusion coefficient ($D_{oo}$) and gallium vacancy diffusion coefficient for  error function model and reaction-diffusion (R-D) linear model for SLs grown on Sn and Fe doped (010) $\beta\text{-}Ga_20_3$ and annealed at $1000$, $1050$ and $1100^\circ C$.  In the Fe a uniform initial $V_{Ga}$ profile is assumed so now value for $D_VGa$ is reported.  $D_{oo}$ is given assuming indicated initial values for $[V_{Ga}]$ in SL and substrate as explained in Sec. \ref{sec:AlinSn}.}
\begin{ruledtabular}
\begin{tabular*}{\textwidth}{l l p{0.1\linewidth} p{0.09\linewidth} l l p{0.1\linewidth} p{0.12\linewidth} l}
Sample & Dopant & Diffusion Model & Anneal Temp ($^\circ C$) & $D_{oo} (cm^5/s)$ & $D_{V_{Ga}} (cm^2/s)$    &Initial $[V_{Ga}]$ in SL ($cm^{-3}$) & Initial $[V_{Ga}]$ in Substrate ($cm^{-3}$) & GoF\\
\hline
SnA	&Sn	&Erf	&1000	&$7.0\pm0.5x10^{-31}$	&$6\pm 1x10^{-14}$	&0 & $10^{16}$  &0.01185	\\
SnA	&Sn	&Linear	&1000	&$6.0\pm0.5x10^{-31}$	&$4\pm1x10^{-13}$	&0 & $10^{16}$   &0.01542	\\
SnB	&Sn	&Erf	&1050	&$2.7\pm0.1x10^{-30}$	&$6\pm1x10^{-14}$	&0 & $10^{16}$  &0.00956	\\
SnB	&Sn	&Linear	&1050	&$2.7\pm0.2x10^{-30}$	&$4\pm0.3x10^{-13}$	&0 & $10^{16}$ &0.00528	\\
SnC	&Sn	&Erf	&1100	&$8.0\pm1.0x10^{-30}$	&$4\pm1x10^{-14}$	&$2.0x10^{13}$ & $2x10^{15}$  &0.01244	\\
SnC	&Sn	&Linear	&1100	&$8.0\pm1.0x10^{-30}$	&$1.4\pm0.2x10^{-12}$	&0 & $2x10^{15}$   &0.01953	\\
Fe	&Fe	&NA	&1050	&$3.5\pm0.5x10^{-32}$	&N/A	&0 & $10^{16}$ &0.05188	\\
\end{tabular*}
\end{ruledtabular}
\end{table*}

In Ref. \cite{frodason_diffusion_2023} (and in some similar intermediate Sn d-SIMS profiles obtained in this work but not shown herein) Sn appears on a log scale to diffuse as a nearly-rigid step with maximum slope set by the temperature but not time.  In the Supplemental Materials we develop an analytical approximation for the x and t evolution of both [$Sn_{Ga}$] and [$V_{Ga}$] within for this R-D process during isothermal annealing.  Remarkably, this is simply a piece-wise-linear function which is initially a step function but in which the initially vertical section pivots vs. time such that its intercepts with the max and min values far away front the interface move as $\sqrt{Dt}$ (see Eq. \ref{eq:Triangle}).  This piece-wise-linear function accurately captures most of the dynamic range of the $Sn_{Ga}-V{Ga}$ steps from Ref. \cite{frodason_diffusion_2023} and our re-analysis of those data.  It only fails to capture some curvature at low concentrations at the diffusion front leading edge and at high concentrations in the substrate.  Since Al diffusion scales with $[V_{Ga}]$, the details of the leading edge are not dominant.  The difference between the classical $erf()$ solution may be seen by comparing the dashed green lines in Fig. \ref{fig:isochronal} d) and e).  Essentially the best-fit model for Al diffusion assuming $erf()$ diffusion for $[V_{Ga}]$ suggests a significant gradient of $[V_{Ga}]$ (and thus $Sn_{Ga}$) should remain at the end of our experiments, while the RD model and our approximation predict the $[V_{Ga}]$ and [Sn] should be closer to uniform at the end of the experiment.  The later is in agreement with the measured, nearly-uniform final [Sn] profiles as exemplified by Fig. \ref{fig:SnvFeSchematic}b).  Additionally, using the piecewise-linear RD approximation yields an order of magnitude better GOF values and allows us to recover Arrhenius diffusion behavior for both Al and $V_{Ga}$.  We take these facts to be strong indicators that the piecewise RD approximation for $V_{Ga}$ diffusion more accurately reflects the actual $V_{Ga}$ diffusion in these SLs.  The magnitude of of $D_{V_{Ga}}$ extracted from under the RD model show remarkable agreement with $D_{Sn}$  in \cite{frodason_diffusion_2023} despite $D_{V_{Ga}}$ being extracted from our [Al] profiles and $D_{Sn}$ in theirs from [Sn] profiles, indicating that in both cases the diffusivity of $Sn_{Ga}V_{Ga}$ complexes or at least the interactions between $Sn_{Ga}$ and $V_{Ga}$ determine the is the rate limiting step(s).  

The lack of evidence for $[V_{Ga}]$ in-diffusion from the free surface in our [Al] data herein raises a question for future investigations: what sources, sinks, and transport processes for $[V_{Ga}]$ operate in order to establish equilibration with local Fermi levels and $p_{O_2}$ during annealing and growth?  The creation and transport of vacancies bears further examination in $Ga_2O_3$ and in other single crystal diffusion experiments.  However, for the immediate purpose of analyzing our Al diffusion data, this question is immaterial since the RD models predict the free $[V_{Ga}]$ able to facilitate Al diffusion to have functional shape closely mimicking the Sn profile.

The Supplemental Materials contain a thorough discussion of whether or not $V_{Ga}$ formation in the bulk or transport from the free surface (where formation presumably has lower activation barrier) is detectable in our experiments and those of Ref. \cite{frodason_diffusion_2023}. Succinctly, the lower extent of diffusion of Al close to the free surface appears to rule out large concentrations of $[V_{Ga}]$ being injected from that surface (“large” here being relative to ~at\% concentrations of Al).  Also, our models account for the Al diffusion without requiring generation of $V_{Ga}$ in the SLs, only the diffusion of those initially present in the substrate.  In Ref. \cite{frodason_diffusion_2023} such mechanisms are implicit requirements of the assumption that $[V_{Ga}]$ equilibrates with the local Fermi level.  Since Sn in \cite{frodason_diffusion_2023} concentrations range from $10^{14}$ to $10^{18} /cm^3$, these experiments are sensitive to much lower concentrations of $V_{Ga}$.  Thus we can only conclude that any bulk generation or injection of $V_{Ga}$ from the free surface must be below the capability of the present experiments with Al to resolve and further experiments with smaller tracer concentrations, longer times, or higher temperatures should be used to elucidate the kinetics of $V_{Ga}$ introduction. 

From a visual comparison alone, the sample annealed at $1050^\circ C$ (Fig \ref{fig:isochronal}b)) shows a greater extent of diffusion than the one annealed at $1100^\circ C$ (Fig \ref{fig:isochronal}c)), counter to what would be expected for Arrhenius activation if the two samples had identical initial concentrations of $[V_{Ga}]$.  

This was initially a source of great consternation until we realized that the sample annealed at $1100^\circ C$ (Fig \ref{fig:isochronal}c) ) was grown in an earlier growth run using a different substrate, while the samples annealed at $1000$ and $1050^\circ C$ (Fig \ref{fig:isochronal}a,b) ) were grown at a later date on pieces of a new  Sn-doped (010) $\beta\text{-}Ga_2O_3$ wafer different than the one for the $1100^\circ C$ one.  $[V_{Ga}]$ is likely to vary spatially within EFG-grown bulk crystals, thus variation from wafer to wafer may occur.  Likewise, despite using nominally the same growth recipe for all three samples, some variation is reasonably likely in $[V_{Ga}]$ within the SL between growth runs.  The finite difference model in Sec. \ref{sec:methods} ultimately depends on $D_{Al}(x,t)$ as defined in Eq. \ref{eq:Doo}; thus fitting via simulation does not directly measure $[V_{Ga}]$.  Thus in our final analysis, we assumed $[V_{Ga}]=10^16 /cm^3$ in the substrates for the $1000$ and $1050^\circ C$ annealed samples, but explored possibilities that the substrate for the $1000^\circ C$ sample had a different $[V_{Ga}]$.  In all cases, the value of $[V_{Ga}]$ in the SL was determined independently for each sample through the fitting process.    After testing a wide range of scenarios, the smaller value of $D_{Al} = D_{oo}\cdot[V_{Ga}]$ for the sample annealed at $1100 ^\circ C$ is best explained by lower initial $[V_{Ga}]$ in that sample’s substrate compared to the samples (from a different OMVPE growth and cut from a different wafer) annealed at $1000$ and $1050^\circ C$.    After accounting for this difference in $[V_{Ga}]$, and  rerunning simulations with initial $[V_{Ga}]$ in the substrate reduced from $10^{16} /cm^3$ to $2x10^{15} /cm^3$,  more sensible behavior for $D_{oo}(T)$ compatible with an Arrhenius law with activation energy $3.9\pm0.3$ eV is obtained.  This emphasizes the extreme sensitivity to the local $[V_{Ga}]$ concentration for self- and impurity-diffusion, whilst also demonstrating the efficacy of our modeling at extracting physically-reasonable behavior even from samples having heterogeneous initial conditions.  The 3.9 eV activation energy for $D_{oo}(T)$ is comparable to the value of 4.2 eV obtained for $Al_{Ga}$ diffusion in (-201) $\beta{\text-}Ga_2O_3$ in ref \cite{uhlendorf_oxygen_2023}.  Table  \ref{tab:resultssummary} summarizes all of the parameters from best-fit scenarios under both the $erf()$ and piece-wise-linear RD models for the $V_{Ga}$ diffusion (our final analysis concludes that the later is the correct model).

In Fig. \ref{fig:SnvFeSchematic}a,b) we compare two SLs annealed at $1050^\circ C$ on Fe and Sn doped substrates.  On the Sn doped substrate, the best-fit scenario within the piece-wise-linear model of $V_{Ga}$ diffusion yields $D_{oo}=2.7x10^{-30} cm^{5}/s$ at an initial value of $[V_{Ga}] =10^{16} /cm^3$ in the substrate.  $D_{oo}$ for Al in the absence of $V_{Ga}$ complexing with Sn would be expected to be the same or larger, depending on how significantly the presence of Sn limits the ability of $V_{Ga}$ to mediate Al diffusion.  If we do assume that $D_{oo}$ is the same for samples grown on both Fe- or Sn-doped substrates, we estimate the initial $[V_{Ga}]$  in the substrate for samples grown on Fe-doped substrates would be $10^{14} /cm^3$, o r $\approx 100$ times lower that for the Sn-doped substrates.  This is on the lower end of the range of estimated ratios based on expected Fermi level positions, but given the lack of direct comparison data and uncertainty regarding the effects of Sn and Fe co-diffusing with $V_{Ga}$, considered reasonable.

Returning to the issue of vacancy conservation, our data and analyses point to these experiments lasting on the order of a day or more at $1000-1100^\circ C$ $V_{Ga}$ diffusion occurring over a distance of 1.5-2.5 um within a transient, not equilibrated regime.    The data show no indication for in-diffusion from the surface, the density of vacancy sources like dislocations is small ($10^4-10^5 /cm^2$), spontaneous vacancy pair formation has a large activation energy.  These are interesting and impactful results as most analyses of similar diffusion experiments assume that mediating native defects equilibrate quickly compared to impurity or tracer diffusion timescales.  Herein, our evidence indicates cation vacancies remaining in a transient diffusion regime over the 1.5-2.5 $\mu m$ SL thickness for most or all of experiments at $1000-1100 ^\circ C$ typically lasting about a full day (at least for concentrations necessary for diffusion of Al at $10^{18}-10^{21} /cm^3$ concentrations).  This may be a unique experimental case in that the homoepitaxial samples have low densities of vacancy sources/sinks (c.f. grain boundaries or dislocations) and our diffusing species were not introduced as reactive surface layers nor were they ion implanted (both of which can introduce or consume mediating native defects).  Higher temperatures and different chemical boundary conditions at free surfaces may change this, but these experiments provide a valuable existence proof of cases where this typically foundational assumption of analysis of native defect-mediated diffusion is not fulfilled.

\section{\label{sec:survey}Diffusion of Unintentional Impurities} 

\begin{figure}
    \centering
    \includegraphics[width=\columnwidth]{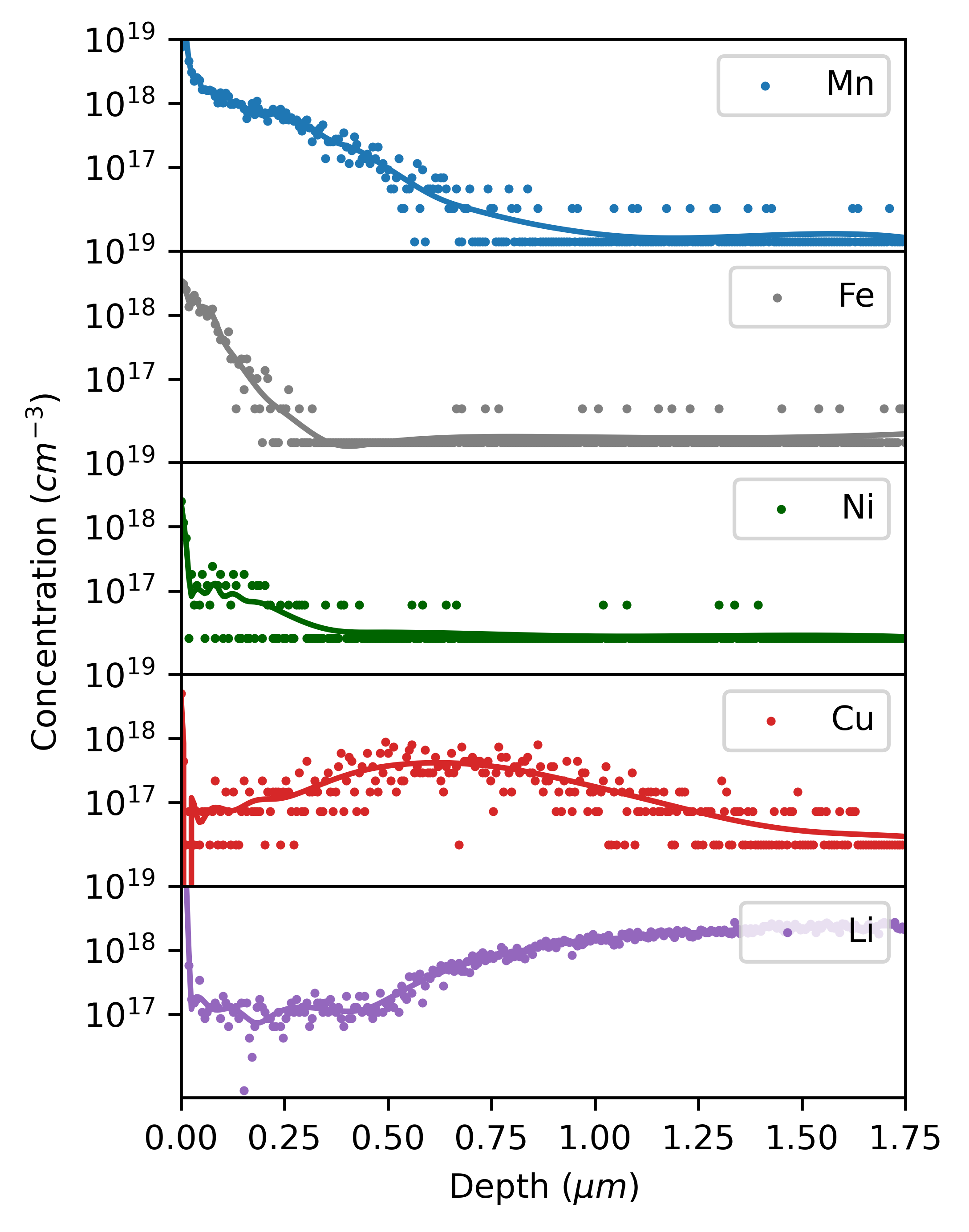}
    \caption{Low spatial and concentration resolution “survey” SIMS profiles of five species which show high concentration and non-homogenous behavior for SL on Sn doped substrate annealed at $1000^\circ C$ for 30 hours in oxygen.  Each survey is fit with a univariate spline function (solid line).}
    \label{fig:survey}
\end{figure}

In addition to the d-SIMS of Al, Fe, and Sn presented so far, we also used TOF-SIMS 3D reconstructions and survey d-SIMS detecting all elements present (at lower spatial and concentration resolutions).  The revealed that many other   elements besides Fe or Sn diffused out of a Sn-doped substrate and through the SLs.  Figure \ref{fig:survey} shows concentration-calibrated SIMS profiles for five illustrative metals present in the SL shown in Fig. \ref{fig:isochronal}b), annealed at $1000 ^\circ C$.  After collecting the 20 hour anneal SIMS profile shown in Fig. \ref{fig:isochronal}b), the sample  was immersed in aqua regia for 10 minutes to remove any possibility of metal contamination at the surface, and then annealed in $O_2$ for an additional 10 hours (total 30 hours) as part of efforts to eliminate possibilities of Fe contamination depositing on the surface and then diffusing into the SL while annealing inside the tube furnace (Supplemental Materials).    Given the known presence of many of these elements in wafers grown from melts held in Ir crucibles- e.g. EFG\cite{mccloy_growth_2022,kuramata_high-quality_2016} and the high purity of OMVPE-grown layers, we infer that these elements diffused into the SLs from the substrate.   

Mn, Fe and Ni accumulate at the surface at concentrations, and to depths, which are unlikely to be due to artifacts.  [Cu] exhibits a concentration peak around the middle of the superlattice structure, perhaps indicating interactions between its charge states and the Fermi level within a space charge region.  All of these transition metals are likely capable of multiple interstitial and substitutional diffusion modes in multiple charge states.  Ti was also detected at $10^{16} /cm^3$ uniform in depth through the d-SIMS survey scan, while Si and Na were also detected in TOF SIMS.  Li being extremely small is suspected to be capable of fast interstitial diffusion and is depleted rather than enriched close to the surface.  These different behaviors are consistent with an electro-chemical potential gradient affecting different elements and charge states differently near the surface.  A depletion width and built-in potential $\phi(x)$ induced by surface Fermi-level pinning is the simplest example and defects with different relative charges q would accumulate or deplete as $exp\left(\frac{q\phi\left(x\right)}{k_BT}\right)$ self-consistently following the Poisson-Boltzmann equation.  However, it has been frequently observed that $\beta\text{-}Ga_2O_3$ surfaces, especially in the presence of additional impurities, can change to the different $\gamma\text{-}Ga_2O_3$ phase which would of course change the chemical potential for each impurity, perhaps causing them to accumulate.  Further analysis of these profiles is beyond the scope of this work, but the diffusion of these impurities, some like Cu and Li being notoriously fast diffusers in semiconductors, calls for further quantification of impurities and their diffusion in multiple phases of $Ga_2O_3$ and in devices from fabrication to long-term degradation under high fields and temperatures.

\section{Contextualizing Results} \label{sec:contextualizing}

\begin{figure*}
    \centering
    \includegraphics[width=\textwidth]{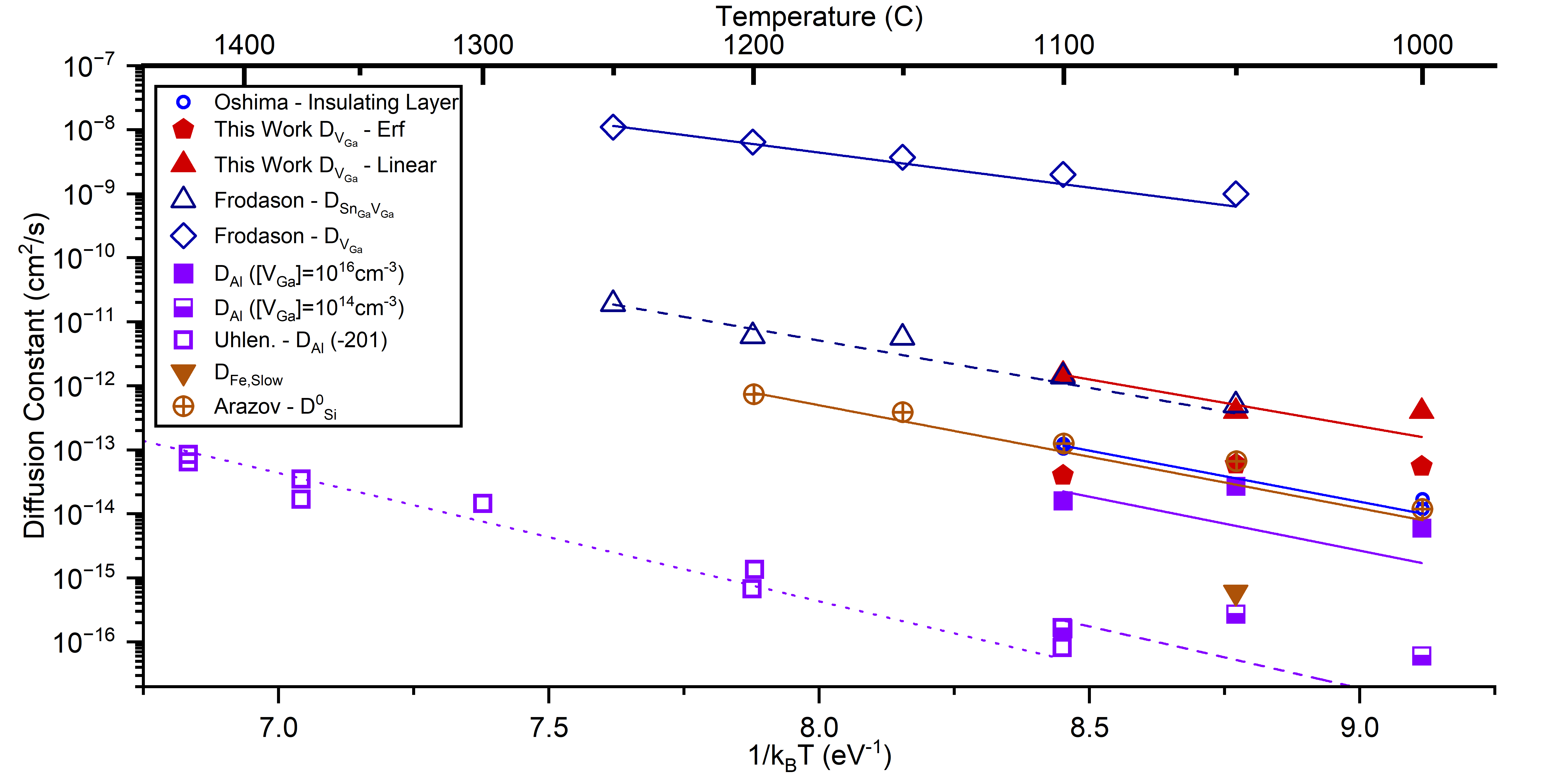}
    \caption{Arrhenius plot for the extent of formation of semi-insulating layer by Oshima et al.\cite{oshima_formation_2013} (blue circle), the assumed $D_{V_{Ga}}$ and $D_{Sn_{Ga}V_{Ga}}$ extracted from best fit in work by Frodason et al\cite{frodason_diffusion_2023} (blue rhombus and triangle respectively), and from this work: $D_{V_{Ga}}$ extracted from the Al tracer diffusion under free, error function diffusion model and linear reaction diffusion model (red pentagons and red triangles respectively).  Based on our extracted values of $D_{oo}(T)$ for Al, we present $D_{Al}$ for two concentrations of $V_{Ga}$ (see Eq. \ref{eq:Doo}): $[VGa]=10^{16} cm^{-3}$ appropriate for Al in Sn-doped wafers (solid purple squares) and $[VGa]=10^{14} cm^{-3}$ appropriate for semi-insulating or oxygen annealed samples (half filled purple square).  The later agrees well with recently published data by Uhlendorf and Schmitt \cite{uhlendorf_oxygen_2023}, further strengthening the case for validity of our values.  $D_{Fe}$ extracted from the slow-diffusing component in Fig \ref{fig:SnvFeSchematic}a, is shown as inverted orange triangles.  In addition, $D_{Si}^0$ from \cite{azarov_activation_2021} is shown as brown crossed-circles; this follows a similar activation energy but is apparently slightly slower than the diffusion of $(Sn_{Ga}-V_{Ga})$ complexes.}
    \label{fig:MasterPlot}
\end{figure*}

Figure \ref{fig:MasterPlot} compares measured and computed diffusivities in $Ga_2O_3$ from this work and that from literature.  Red pentagons and triangles give the diffusivities inferred in this work for $V_{Ga}$ assuming no interaction with Sn (erf) and coupled reaction-diffusion with Sn (piece-wise linear).  $D_{V_{Ga}}$ extracted from our Al diffusion data over the investigated temperature range assuming only the piece-wise linear shape of the $[V_{Ga}](x,t)$ evolution (making no assumptions of diffusion constant) have remarkable agreement with the diffusion of $Sn_{Ga}-V_{Ga}$ complexes reported by Frodason\cite{frodason_diffusion_2023} and follow a sensible Arrhenius trend with activation energy $1.86\pm 1.13 eV$  The values extracted from best-fits at different temperatures assuming free diffusion ($erf(x,t)$) are an order of magnitude lower and deviate from Arrhenius behavior.  Blue rhombuses plot the modeled values of free diffusion of uncomplexed $D_{V_{Ga}}$ from Ref \cite{frodason_diffusion_2023}.  It is clear that interactions of acceptor-like $V_Ga$ especially with donor-like defects significantly slows their diffusion. 

Since Al diffusion is proportional to $[V_{Ga}]$ and this varies with depth and time in the SLs on Sn-doped substrates, we present $D_{Al}$  extracted for each temperature, assuming a uniform concentration local $[V_{Ga}]=10^{16}/cm^3$.  To emphasize the dependence of $D_{Al}$ on local $[V_{Ga}]$ we also give a second dataset for $D_{Al}$ assuming a uniform local $[VGa] =10^{14} /cm^3$.  This latter set gives good agreement in magnitude with $D_{Al}$ derived from the interdiffusion of a thin $Al_2O_3$ layer on $Ga_2O_3$ during oxygen annealing\cite{uhlendorf_oxygen_2023}.  The data shown as an inverted brown triangle is the value for the extracted diffusion constants for the slow diffusion channel of Fe out of an Fe-doped substrate.  This value is similar to those experienced by Al, so we speculate may be vacancy- mediated but modified by the expected Coulomb repulsion between $V_{Ga}$ and $Fe_{Ga}$ acceptors compared to the case of Al.  Further detailed studies will be needed to understand the diffusion of Fe and other metals prone to charge state changes and possibly diffusing by multiple mechanisms simultaneously.  

Oshima et al., used capacitance methods to measure the expansion of an insulating layer on the surface of n-type samples up to a few $\mu m$ thickness during $O_2$ annealing.  The kinetics of layer thickening were consistent with a diffusion-limited process, thus we plot their equivalent diffusion constant as blue circles.  It is very interesting to note that these extracted diffusion constants are much closer to the values for impurities like Al or Fe diffusing to the surface in our samples than they are to any of the cation vacancy diffusion constants or oxygen tracer diffusion constants recently reported.  Thus, we speculate that at least part of the phenomenology of insulating layers forming on the surface of n-type $Ga_2O_3$ crystals may be caused by out-diffusion of unintentional cation impurities towards the O-rich boundary condition, possible surface band bending, and possibly phase changes to g-Ga2O3 with impurity pile-up, rather than directly by in-diffusion of O or Ga native defects as the simplest models might assume.  Clearly, there are many questions to be explored with direct experiment and theory in the diffusion and annealing behavior of native defects and impurities in $Ga_2O_3$.  Our results point out many places where unchallenged assumptions regarding native defect processes in diffusion experiments should be revisited and directly tested.  

\section{Conclusions}

While many papers have shown a correlation between gallium vacancies and n-type carrier concentration after oxygen annealing, the exact dynamics of this process have not previously been explored.  We have shown via Al tracer diffusion, that $V_{Ga}$ are available in great numbers in Sn doped substrates, diffusing outward towards the epi-layers in our SL experiments.  However, the rate of diffusion is much faster than that indicated for the formation of the semi-insulating layer measured by Oshima.  We developed a modified finite difference scheme in which the diffusion of impurities such as Al is dependent on the concentration of $V_{Ga}$, which is also diffusing.  The finite differences model herein developed shows no evidence of Ga vacancies being introduced from the surface, though these may be introduced at lower concentrations.  
The measured diffusion rates extracted for $V_{Ga}$ are far lower than those predicted for free vacancies based on DFT calculations.  Instead they agree well with the measured rates of Sn diffusion, indicating that the formation and diffusion of defect complexes is the primary rate controlling step in the diffusion of the charge-neutral Al defects.  
The study of unintentionally introduced impurities like Fe, Mn, Cu must be explored further, and may yet play an important role in elucidating the formation of the semi-insulating layer.  The accumulation of transition metals at the surface suggests that the effect of the surface potential must be taken into account in future work.

\begin{acknowledgments}
This work was funded by the Air Force Office of Scientific Research – Multidisciplinary Research Program of the University Research Initiative (MURI) – AFOSR award number FA9550-18-1-0507.  We thank EAG Eurofins for their assistance in SIMS analysis.   We also thank Northrop Grumman-Synoptics for the supply of $\beta\text{-}Ga_2O_3$ boule pieces.  We thank Dr. Klaus Magnus Johansen and Dr. Ymir K. Frodason for the use of their data and modeling of $Sn_{Ga}$ and $V_{Ga}$ reaction-diffusion.  We thank Prof. Danny Feezell and his group at UNM for high-resolution Xray diffraction scans on early samples to help characterize the growth rate variations.

\end{acknowledgments}

\section*{Data Availability Statement}

The data that support the findings of this study are available within the article [and its supplementary material].
%\nocite{*}
\bibliography{SnSLPaper,SnSLPaper22Feb}% Produces the bibliography via BibTeX.

%merlin.mbs aipnum4-1.bst 2010-07-25 4.21a (PWD, AO, DPC) hacked
%Control: key (0)
%Control: author (8) initials jnrlst
%Control: editor formatted (1) identically to author
%Control: production of article title (0) allowed
%Control: page (1) range
%Control: year (1) truncated
%Control: production of eprint (0) enabled
\begin{thebibliography}{41}%
\makeatletter
\providecommand \@ifxundefined [1]{%
 \@ifx{#1\undefined}
}%
\providecommand \@ifnum [1]{%
 \ifnum #1\expandafter \@firstoftwo
 \else \expandafter \@secondoftwo
 \fi
}%
\providecommand \@ifx [1]{%
 \ifx #1\expandafter \@firstoftwo
 \else \expandafter \@secondoftwo
 \fi
}%
\providecommand \natexlab [1]{#1}%
\providecommand \enquote  [1]{``#1''}%
\providecommand \bibnamefont  [1]{#1}%
\providecommand \bibfnamefont [1]{#1}%
\providecommand \citenamefont [1]{#1}%
\providecommand \href@noop [0]{\@secondoftwo}%
\providecommand \href [0]{\begingroup \@sanitize@url \@href}%
\providecommand \@href[1]{\@@startlink{#1}\@@href}%
\providecommand \@@href[1]{\endgroup#1\@@endlink}%
\providecommand \@sanitize@url [0]{\catcode `\\12\catcode `\$12\catcode `\&12\catcode `\#12\catcode `\^12\catcode `\_12\catcode `\%12\relax}%
\providecommand \@@startlink[1]{}%
\providecommand \@@endlink[0]{}%
\providecommand \url  [0]{\begingroup\@sanitize@url \@url }%
\providecommand \@url [1]{\endgroup\@href {#1}{\urlprefix }}%
\providecommand \urlprefix  [0]{URL }%
\providecommand \Eprint [0]{\href }%
\providecommand \doibase [0]{http://dx.doi.org/}%
\providecommand \selectlanguage [0]{\@gobble}%
\providecommand \bibinfo  [0]{\@secondoftwo}%
\providecommand \bibfield  [0]{\@secondoftwo}%
\providecommand \translation [1]{[#1]}%
\providecommand \BibitemOpen [0]{}%
\providecommand \bibitemStop [0]{}%
\providecommand \bibitemNoStop [0]{.\EOS\space}%
\providecommand \EOS [0]{\spacefactor3000\relax}%
\providecommand \BibitemShut  [1]{\csname bibitem#1\endcsname}%
\let\auto@bib@innerbib\@empty
%</preamble>
\bibitem [{\citenamefont {Pearton}\ \emph {et~al.}()\citenamefont {Pearton}, \citenamefont {Yang}, \citenamefont {Cary}, \citenamefont {Ren}, \citenamefont {Kim}, \citenamefont {Tadjer},\ and\ \citenamefont {Mastro}}]{pearton_review_2018}%
  \BibitemOpen
  \bibfield  {author} {\bibinfo {author} {\bibfnamefont {S.~J.}\ \bibnamefont {Pearton}}, \bibinfo {author} {\bibfnamefont {J.}~\bibnamefont {Yang}}, \bibinfo {author} {\bibfnamefont {P.~H.}\ \bibnamefont {Cary}}, \bibinfo {author} {\bibfnamefont {F.}~\bibnamefont {Ren}}, \bibinfo {author} {\bibfnamefont {J.}~\bibnamefont {Kim}}, \bibinfo {author} {\bibfnamefont {M.~J.}\ \bibnamefont {Tadjer}}, \ and\ \bibinfo {author} {\bibfnamefont {M.~A.}\ \bibnamefont {Mastro}},\ }\bibfield  {title} {\enquote {\bibinfo {title} {A review of ga2o3 materials, processing, and devices},}\ }\href {\doibase 10.1063/1.5006941} {\ \textbf {\bibinfo {volume} {5}},\ \bibinfo {pages} {011301}},\ \bibinfo {note} {publisher: American Institute of Physics}\BibitemShut {NoStop}%
\bibitem [{\citenamefont {Green}\ \emph {et~al.}()\citenamefont {Green}, \citenamefont {Speck}, \citenamefont {Xing}, \citenamefont {Moens}, \citenamefont {Allerstam}, \citenamefont {Gumaelius}, \citenamefont {Neyer}, \citenamefont {Arias-Purdue}, \citenamefont {Mehrotra}, \citenamefont {Kuramata}, \citenamefont {Sasaki}, \citenamefont {Watanabe}, \citenamefont {Koshi}, \citenamefont {Blevins}, \citenamefont {Bierwagen}, \citenamefont {Krishnamoorthy}, \citenamefont {Leedy}, \citenamefont {Arehart}, \citenamefont {Neal}, \citenamefont {Mou}, \citenamefont {Ringel}, \citenamefont {Kumar}, \citenamefont {Sharma}, \citenamefont {Ghosh}, \citenamefont {Singisetti}, \citenamefont {Li}, \citenamefont {Chabak}, \citenamefont {Liddy}, \citenamefont {Islam}, \citenamefont {Rajan}, \citenamefont {Graham}, \citenamefont {Choi}, \citenamefont {Cheng},\ and\ \citenamefont {Higashiwaki}}]{green_-gallium_2022}%
  \BibitemOpen
  \bibfield  {author} {\bibinfo {author} {\bibfnamefont {A.~J.}\ \bibnamefont {Green}}, \bibinfo {author} {\bibfnamefont {J.}~\bibnamefont {Speck}}, \bibinfo {author} {\bibfnamefont {G.}~\bibnamefont {Xing}}, \bibinfo {author} {\bibfnamefont {P.}~\bibnamefont {Moens}}, \bibinfo {author} {\bibfnamefont {F.}~\bibnamefont {Allerstam}}, \bibinfo {author} {\bibfnamefont {K.}~\bibnamefont {Gumaelius}}, \bibinfo {author} {\bibfnamefont {T.}~\bibnamefont {Neyer}}, \bibinfo {author} {\bibfnamefont {A.}~\bibnamefont {Arias-Purdue}}, \bibinfo {author} {\bibfnamefont {V.}~\bibnamefont {Mehrotra}}, \bibinfo {author} {\bibfnamefont {A.}~\bibnamefont {Kuramata}}, \bibinfo {author} {\bibfnamefont {K.}~\bibnamefont {Sasaki}}, \bibinfo {author} {\bibfnamefont {S.}~\bibnamefont {Watanabe}}, \bibinfo {author} {\bibfnamefont {K.}~\bibnamefont {Koshi}}, \bibinfo {author} {\bibfnamefont {J.}~\bibnamefont {Blevins}}, \bibinfo {author} {\bibfnamefont {O.}~\bibnamefont {Bierwagen}}, \bibinfo {author} {\bibfnamefont {S.}~\bibnamefont
  {Krishnamoorthy}}, \bibinfo {author} {\bibfnamefont {K.}~\bibnamefont {Leedy}}, \bibinfo {author} {\bibfnamefont {A.~R.}\ \bibnamefont {Arehart}}, \bibinfo {author} {\bibfnamefont {A.~T.}\ \bibnamefont {Neal}}, \bibinfo {author} {\bibfnamefont {S.}~\bibnamefont {Mou}}, \bibinfo {author} {\bibfnamefont {S.~A.}\ \bibnamefont {Ringel}}, \bibinfo {author} {\bibfnamefont {A.}~\bibnamefont {Kumar}}, \bibinfo {author} {\bibfnamefont {A.}~\bibnamefont {Sharma}}, \bibinfo {author} {\bibfnamefont {K.}~\bibnamefont {Ghosh}}, \bibinfo {author} {\bibfnamefont {U.}~\bibnamefont {Singisetti}}, \bibinfo {author} {\bibfnamefont {W.}~\bibnamefont {Li}}, \bibinfo {author} {\bibfnamefont {K.}~\bibnamefont {Chabak}}, \bibinfo {author} {\bibfnamefont {K.}~\bibnamefont {Liddy}}, \bibinfo {author} {\bibfnamefont {A.}~\bibnamefont {Islam}}, \bibinfo {author} {\bibfnamefont {S.}~\bibnamefont {Rajan}}, \bibinfo {author} {\bibfnamefont {S.}~\bibnamefont {Graham}}, \bibinfo {author} {\bibfnamefont {S.}~\bibnamefont {Choi}}, \bibinfo
  {author} {\bibfnamefont {Z.}~\bibnamefont {Cheng}}, \ and\ \bibinfo {author} {\bibfnamefont {M.}~\bibnamefont {Higashiwaki}},\ }\bibfield  {title} {\enquote {\bibinfo {title} {β-gallium oxide power electronics},}\ }\href {\doibase 10.1063/5.0060327} {\ \textbf {\bibinfo {volume} {10}},\ \bibinfo {pages} {029201}}\BibitemShut {NoStop}%
\bibitem [{\citenamefont {Tsao}\ \emph {et~al.}()\citenamefont {Tsao}, \citenamefont {Chowdhury}, \citenamefont {Hollis}, \citenamefont {Jena}, \citenamefont {Johnson}, \citenamefont {Jones}, \citenamefont {Kaplar}, \citenamefont {Rajan}, \citenamefont {Van~de Walle}, \citenamefont {Bellotti}, \citenamefont {Chua}, \citenamefont {Collazo}, \citenamefont {Coltrin}, \citenamefont {Cooper}, \citenamefont {Evans}, \citenamefont {Graham}, \citenamefont {Grotjohn}, \citenamefont {Heller}, \citenamefont {Higashiwaki}, \citenamefont {Islam}, \citenamefont {Juodawlkis}, \citenamefont {Khan}, \citenamefont {Koehler}, \citenamefont {Leach}, \citenamefont {Mishra}, \citenamefont {Nemanich}, \citenamefont {Pilawa-Podgurski}, \citenamefont {Shealy}, \citenamefont {Sitar}, \citenamefont {Tadjer}, \citenamefont {Witulski}, \citenamefont {Wraback},\ and\ \citenamefont {Simmons}}]{tsao_ultrawide-bandgap_2018}%
  \BibitemOpen
  \bibfield  {author} {\bibinfo {author} {\bibfnamefont {J.~Y.}\ \bibnamefont {Tsao}}, \bibinfo {author} {\bibfnamefont {S.}~\bibnamefont {Chowdhury}}, \bibinfo {author} {\bibfnamefont {M.~A.}\ \bibnamefont {Hollis}}, \bibinfo {author} {\bibfnamefont {D.}~\bibnamefont {Jena}}, \bibinfo {author} {\bibfnamefont {N.~M.}\ \bibnamefont {Johnson}}, \bibinfo {author} {\bibfnamefont {K.~A.}\ \bibnamefont {Jones}}, \bibinfo {author} {\bibfnamefont {R.~J.}\ \bibnamefont {Kaplar}}, \bibinfo {author} {\bibfnamefont {S.}~\bibnamefont {Rajan}}, \bibinfo {author} {\bibfnamefont {C.~G.}\ \bibnamefont {Van~de Walle}}, \bibinfo {author} {\bibfnamefont {E.}~\bibnamefont {Bellotti}}, \bibinfo {author} {\bibfnamefont {C.~L.}\ \bibnamefont {Chua}}, \bibinfo {author} {\bibfnamefont {R.}~\bibnamefont {Collazo}}, \bibinfo {author} {\bibfnamefont {M.~E.}\ \bibnamefont {Coltrin}}, \bibinfo {author} {\bibfnamefont {J.~A.}\ \bibnamefont {Cooper}}, \bibinfo {author} {\bibfnamefont {K.~R.}\ \bibnamefont {Evans}}, \bibinfo {author}
  {\bibfnamefont {S.}~\bibnamefont {Graham}}, \bibinfo {author} {\bibfnamefont {T.~A.}\ \bibnamefont {Grotjohn}}, \bibinfo {author} {\bibfnamefont {E.~R.}\ \bibnamefont {Heller}}, \bibinfo {author} {\bibfnamefont {M.}~\bibnamefont {Higashiwaki}}, \bibinfo {author} {\bibfnamefont {M.~S.}\ \bibnamefont {Islam}}, \bibinfo {author} {\bibfnamefont {P.~W.}\ \bibnamefont {Juodawlkis}}, \bibinfo {author} {\bibfnamefont {M.~A.}\ \bibnamefont {Khan}}, \bibinfo {author} {\bibfnamefont {A.~D.}\ \bibnamefont {Koehler}}, \bibinfo {author} {\bibfnamefont {J.~H.}\ \bibnamefont {Leach}}, \bibinfo {author} {\bibfnamefont {U.~K.}\ \bibnamefont {Mishra}}, \bibinfo {author} {\bibfnamefont {R.~J.}\ \bibnamefont {Nemanich}}, \bibinfo {author} {\bibfnamefont {R.~C.~N.}\ \bibnamefont {Pilawa-Podgurski}}, \bibinfo {author} {\bibfnamefont {J.~B.}\ \bibnamefont {Shealy}}, \bibinfo {author} {\bibfnamefont {Z.}~\bibnamefont {Sitar}}, \bibinfo {author} {\bibfnamefont {M.~J.}\ \bibnamefont {Tadjer}}, \bibinfo {author} {\bibfnamefont
  {A.~F.}\ \bibnamefont {Witulski}}, \bibinfo {author} {\bibfnamefont {M.}~\bibnamefont {Wraback}}, \ and\ \bibinfo {author} {\bibfnamefont {J.~A.}\ \bibnamefont {Simmons}},\ }\bibfield  {title} {\enquote {\bibinfo {title} {Ultrawide-bandgap semiconductors: Research opportunities and challenges},}\ }\href {\doibase 10.1002/aelm.201600501} {\ \textbf {\bibinfo {volume} {4}},\ \bibinfo {pages} {1600501}},\ \bibinfo {note} {\_eprint: https://onlinelibrary.wiley.com/doi/pdf/10.1002/aelm.201600501}\BibitemShut {NoStop}%
\bibitem [{\citenamefont {Oshima}\ \emph {et~al.}({\natexlab{a}})\citenamefont {Oshima}, \citenamefont {Okuno}, \citenamefont {Arai}, \citenamefont {Suzuki}, \citenamefont {Ohira},\ and\ \citenamefont {Fujita}}]{oshima_vertical_2008}%
  \BibitemOpen
  \bibfield  {author} {\bibinfo {author} {\bibfnamefont {T.}~\bibnamefont {Oshima}}, \bibinfo {author} {\bibfnamefont {T.}~\bibnamefont {Okuno}}, \bibinfo {author} {\bibfnamefont {N.}~\bibnamefont {Arai}}, \bibinfo {author} {\bibfnamefont {N.}~\bibnamefont {Suzuki}}, \bibinfo {author} {\bibfnamefont {S.}~\bibnamefont {Ohira}}, \ and\ \bibinfo {author} {\bibfnamefont {S.}~\bibnamefont {Fujita}},\ }\bibfield  {title} {\enquote {\bibinfo {title} {Vertical solar-blind deep-ultraviolet schottky photodetectors based on $\beta$-ga $_{\textrm{2}}$ o $_{\textrm{3}}$ substrates},}\ }\href {\doibase 10.1143/APEX.1.011202} {\ \textbf {\bibinfo {volume} {1}},\ \bibinfo {pages} {011202} ({\natexlab{a}})}\BibitemShut {NoStop}%
\bibitem [{\citenamefont {Varley}\ \emph {et~al.}()\citenamefont {Varley}, \citenamefont {Weber}, \citenamefont {Janotti},\ and\ \citenamefont {Van~de Walle}}]{varley_oxygen_2010}%
  \BibitemOpen
  \bibfield  {author} {\bibinfo {author} {\bibfnamefont {J.~B.}\ \bibnamefont {Varley}}, \bibinfo {author} {\bibfnamefont {J.~R.}\ \bibnamefont {Weber}}, \bibinfo {author} {\bibfnamefont {A.}~\bibnamefont {Janotti}}, \ and\ \bibinfo {author} {\bibfnamefont {C.~G.}\ \bibnamefont {Van~de Walle}},\ }\bibfield  {title} {\enquote {\bibinfo {title} {Oxygen vacancies and donor impurities in β-ga2o3},}\ }\href {\doibase 10.1063/1.3499306} {\ \textbf {\bibinfo {volume} {97}},\ \bibinfo {pages} {142106}},\ \bibinfo {note} {publisher: American Institute of Physics}\BibitemShut {NoStop}%
\bibitem [{\citenamefont {Oshima}\ \emph {et~al.}({\natexlab{b}})\citenamefont {Oshima}, \citenamefont {Kaminaga}, \citenamefont {Mukai}, \citenamefont {Sasaki}, \citenamefont {Masui}, \citenamefont {Kuramata}, \citenamefont {Yamakoshi}, \citenamefont {Fujita},\ and\ \citenamefont {Ohtomo}}]{oshima_formation_2013}%
  \BibitemOpen
  \bibfield  {author} {\bibinfo {author} {\bibfnamefont {T.}~\bibnamefont {Oshima}}, \bibinfo {author} {\bibfnamefont {K.}~\bibnamefont {Kaminaga}}, \bibinfo {author} {\bibfnamefont {A.}~\bibnamefont {Mukai}}, \bibinfo {author} {\bibfnamefont {K.}~\bibnamefont {Sasaki}}, \bibinfo {author} {\bibfnamefont {T.}~\bibnamefont {Masui}}, \bibinfo {author} {\bibfnamefont {A.}~\bibnamefont {Kuramata}}, \bibinfo {author} {\bibfnamefont {S.}~\bibnamefont {Yamakoshi}}, \bibinfo {author} {\bibfnamefont {S.}~\bibnamefont {Fujita}}, \ and\ \bibinfo {author} {\bibfnamefont {A.}~\bibnamefont {Ohtomo}},\ }\bibfield  {title} {\enquote {\bibinfo {title} {Formation of semi-insulating layers on semiconducting β-ga $_{\textrm{2}}$ o $_{\textrm{3}}$ single crystals by thermal oxidation},}\ }\href {\doibase 10.7567/JJAP.52.051101} {\ \textbf {\bibinfo {volume} {52}},\ \bibinfo {pages} {051101} ({\natexlab{b}})}\BibitemShut {NoStop}%
\bibitem [{\citenamefont {Tuomisto}\ and\ \citenamefont {Makkonen}()}]{tuomisto_defect_2013}%
  \BibitemOpen
  \bibfield  {author} {\bibinfo {author} {\bibfnamefont {F.}~\bibnamefont {Tuomisto}}\ and\ \bibinfo {author} {\bibfnamefont {I.}~\bibnamefont {Makkonen}},\ }\bibfield  {title} {\enquote {\bibinfo {title} {Defect identification in semiconductors with positron annihilation: Experiment and theory},}\ }\href {\doibase 10.1103/RevModPhys.85.1583} {\ \textbf {\bibinfo {volume} {85}},\ \bibinfo {pages} {1583--1631}},\ \bibinfo {note} {publisher: American Physical Society}\BibitemShut {NoStop}%
\bibitem [{\citenamefont {Korhonen}\ \emph {et~al.}()\citenamefont {Korhonen}, \citenamefont {Tuomisto}, \citenamefont {Gogova}, \citenamefont {Wagner}, \citenamefont {Baldini}, \citenamefont {Galazka}, \citenamefont {Schewski},\ and\ \citenamefont {Albrecht}}]{korhonen_electrical_2015}%
  \BibitemOpen
  \bibfield  {author} {\bibinfo {author} {\bibfnamefont {E.}~\bibnamefont {Korhonen}}, \bibinfo {author} {\bibfnamefont {F.}~\bibnamefont {Tuomisto}}, \bibinfo {author} {\bibfnamefont {D.}~\bibnamefont {Gogova}}, \bibinfo {author} {\bibfnamefont {G.}~\bibnamefont {Wagner}}, \bibinfo {author} {\bibfnamefont {M.}~\bibnamefont {Baldini}}, \bibinfo {author} {\bibfnamefont {Z.}~\bibnamefont {Galazka}}, \bibinfo {author} {\bibfnamefont {R.}~\bibnamefont {Schewski}}, \ and\ \bibinfo {author} {\bibfnamefont {M.}~\bibnamefont {Albrecht}},\ }\bibfield  {title} {\enquote {\bibinfo {title} {Electrical compensation by ga vacancies in ga2o3 thin films},}\ }\href {\doibase 10.1063/1.4922814} {\ \textbf {\bibinfo {volume} {106}},\ \bibinfo {pages} {242103}},\ \bibinfo {note} {publisher: American Institute of Physics}\BibitemShut {NoStop}%
\bibitem [{\citenamefont {Ingebrigtsen}\ \emph {et~al.}({\natexlab{a}})\citenamefont {Ingebrigtsen}, \citenamefont {Kuznetsov}, \citenamefont {Svensson}, \citenamefont {Alfieri}, \citenamefont {Mihaila}, \citenamefont {Badstübner}, \citenamefont {Perron}, \citenamefont {Vines},\ and\ \citenamefont {Varley}}]{ingebrigtsen_impact_2019}%
  \BibitemOpen
  \bibfield  {author} {\bibinfo {author} {\bibfnamefont {M.~E.}\ \bibnamefont {Ingebrigtsen}}, \bibinfo {author} {\bibfnamefont {A.~Y.}\ \bibnamefont {Kuznetsov}}, \bibinfo {author} {\bibfnamefont {B.~G.}\ \bibnamefont {Svensson}}, \bibinfo {author} {\bibfnamefont {G.}~\bibnamefont {Alfieri}}, \bibinfo {author} {\bibfnamefont {A.}~\bibnamefont {Mihaila}}, \bibinfo {author} {\bibfnamefont {U.}~\bibnamefont {Badstübner}}, \bibinfo {author} {\bibfnamefont {A.}~\bibnamefont {Perron}}, \bibinfo {author} {\bibfnamefont {L.}~\bibnamefont {Vines}}, \ and\ \bibinfo {author} {\bibfnamefont {J.~B.}\ \bibnamefont {Varley}},\ }\bibfield  {title} {\enquote {\bibinfo {title} {Impact of proton irradiation on conductivity and deep level defects in β-ga2o3},}\ }\href {\doibase 10.1063/1.5054826} {\ \textbf {\bibinfo {volume} {7}},\ \bibinfo {pages} {022510} ({\natexlab{a}})},\ \bibinfo {note} {publisher: American Institute of Physics}\BibitemShut {NoStop}%
\bibitem [{\citenamefont {Jesenovec}\ \emph {et~al.}()\citenamefont {Jesenovec}, \citenamefont {Weber}, \citenamefont {Pansegrau}, \citenamefont {{McCluskey}}, \citenamefont {Lynn},\ and\ \citenamefont {{McCloy}}}]{jesenovec_gallium_2021}%
  \BibitemOpen
  \bibfield  {author} {\bibinfo {author} {\bibfnamefont {J.}~\bibnamefont {Jesenovec}}, \bibinfo {author} {\bibfnamefont {M.~H.}\ \bibnamefont {Weber}}, \bibinfo {author} {\bibfnamefont {C.}~\bibnamefont {Pansegrau}}, \bibinfo {author} {\bibfnamefont {M.~D.}\ \bibnamefont {{McCluskey}}}, \bibinfo {author} {\bibfnamefont {K.~G.}\ \bibnamefont {Lynn}}, \ and\ \bibinfo {author} {\bibfnamefont {J.~S.}\ \bibnamefont {{McCloy}}},\ }\bibfield  {title} {\enquote {\bibinfo {title} {Gallium vacancy formation in oxygen annealed β-ga2o3},}\ }\href {\doibase 10.1063/5.0053325} {\ \textbf {\bibinfo {volume} {129}},\ \bibinfo {pages} {245701}}\BibitemShut {NoStop}%
\bibitem [{\citenamefont {Frodason}\ \emph {et~al.}({\natexlab{a}})\citenamefont {Frodason}, \citenamefont {Zimmermann}, \citenamefont {Verhoeven}, \citenamefont {Weiser}, \citenamefont {Vines},\ and\ \citenamefont {Varley}}]{frodason_multistability_2021}%
  \BibitemOpen
  \bibfield  {author} {\bibinfo {author} {\bibfnamefont {Y.~K.}\ \bibnamefont {Frodason}}, \bibinfo {author} {\bibfnamefont {C.}~\bibnamefont {Zimmermann}}, \bibinfo {author} {\bibfnamefont {E.~F.}\ \bibnamefont {Verhoeven}}, \bibinfo {author} {\bibfnamefont {P.~M.}\ \bibnamefont {Weiser}}, \bibinfo {author} {\bibfnamefont {L.}~\bibnamefont {Vines}}, \ and\ \bibinfo {author} {\bibfnamefont {J.~B.}\ \bibnamefont {Varley}},\ }\bibfield  {title} {\enquote {\bibinfo {title} {Multistability of isolated and hydrogenated ga–o divacancies in β − ga 2 o 3},}\ }\href {\doibase 10.1103/PhysRevMaterials.5.025402} {\ \textbf {\bibinfo {volume} {5}},\ \bibinfo {pages} {025402} ({\natexlab{a}})}\BibitemShut {NoStop}%
\bibitem [{\citenamefont {Karjalainen}\ \emph {et~al.}()\citenamefont {Karjalainen}, \citenamefont {Weiser}, \citenamefont {Makkonen}, \citenamefont {Reinertsen}, \citenamefont {Vines},\ and\ \citenamefont {Tuomisto}}]{karjalainen_interplay_2021}%
  \BibitemOpen
  \bibfield  {author} {\bibinfo {author} {\bibfnamefont {A.}~\bibnamefont {Karjalainen}}, \bibinfo {author} {\bibfnamefont {P.~M.}\ \bibnamefont {Weiser}}, \bibinfo {author} {\bibfnamefont {I.}~\bibnamefont {Makkonen}}, \bibinfo {author} {\bibfnamefont {V.~M.}\ \bibnamefont {Reinertsen}}, \bibinfo {author} {\bibfnamefont {L.}~\bibnamefont {Vines}}, \ and\ \bibinfo {author} {\bibfnamefont {F.}~\bibnamefont {Tuomisto}},\ }\bibfield  {title} {\enquote {\bibinfo {title} {Interplay of vacancies, hydrogen, and electrical compensation in irradiated and annealed \textit{n} -type \textit{β} -ga $_{\textrm{2}}$ o $_{\textrm{3}}$},}\ }\href {\doibase 10.1063/5.0042518} {\ \textbf {\bibinfo {volume} {129}},\ \bibinfo {pages} {165702}}\BibitemShut {NoStop}%
\bibitem [{\citenamefont {Kyrtsos}, \citenamefont {Matsubara},\ and\ \citenamefont {Bellotti}()}]{kyrtsos_migration_2017}%
  \BibitemOpen
  \bibfield  {author} {\bibinfo {author} {\bibfnamefont {A.}~\bibnamefont {Kyrtsos}}, \bibinfo {author} {\bibfnamefont {M.}~\bibnamefont {Matsubara}}, \ and\ \bibinfo {author} {\bibfnamefont {E.}~\bibnamefont {Bellotti}},\ }\bibfield  {title} {\enquote {\bibinfo {title} {Migration mechanisms and diffusion barriers of vacancies in ga 2 o 3},}\ }\href {\doibase 10.1103/PhysRevB.95.245202} {\ \textbf {\bibinfo {volume} {95}},\ \bibinfo {pages} {245202}}\BibitemShut {NoStop}%
\bibitem [{\citenamefont {Sun}\ \emph {et~al.}()\citenamefont {Sun}, \citenamefont {Ooi}, \citenamefont {Ranga}, \citenamefont {Bhattacharyya}, \citenamefont {Krishnamoorthy},\ and\ \citenamefont {Scarpulla}}]{sun_oxygen_2021}%
  \BibitemOpen
  \bibfield  {author} {\bibinfo {author} {\bibfnamefont {R.}~\bibnamefont {Sun}}, \bibinfo {author} {\bibfnamefont {Y.~K.}\ \bibnamefont {Ooi}}, \bibinfo {author} {\bibfnamefont {P.}~\bibnamefont {Ranga}}, \bibinfo {author} {\bibfnamefont {A.}~\bibnamefont {Bhattacharyya}}, \bibinfo {author} {\bibfnamefont {S.}~\bibnamefont {Krishnamoorthy}}, \ and\ \bibinfo {author} {\bibfnamefont {M.~A.}\ \bibnamefont {Scarpulla}},\ }\bibfield  {title} {\enquote {\bibinfo {title} {Oxygen annealing induced changes in defects within β-ga2o3 epitaxial films measured using photoluminescence},}\ }\href {\doibase 10.1088/1361-6463/abdefb} {\ \textbf {\bibinfo {volume} {54}},\ \bibinfo {pages} {174004}},\ \bibinfo {note} {publisher: {IOP} Publishing}\BibitemShut {NoStop}%
\bibitem [{\citenamefont {De~Souza}\ and\ \citenamefont {Kilner}()}]{de_souza_oxygen_1998}%
  \BibitemOpen
  \bibfield  {author} {\bibinfo {author} {\bibfnamefont {R.~A.}\ \bibnamefont {De~Souza}}\ and\ \bibinfo {author} {\bibfnamefont {J.~A.}\ \bibnamefont {Kilner}},\ }\bibfield  {title} {\enquote {\bibinfo {title} {Oxygen transport in la1−{xSrxMn}1−{yCoyO}3±δ perovskites: Part i. oxygen tracer diffusion},}\ }\href {\doibase 10.1016/S0167-2738(97)00499-2} {\ \textbf {\bibinfo {volume} {106}},\ \bibinfo {pages} {175--187}}\BibitemShut {NoStop}%
\bibitem [{\citenamefont {Töpfer}, \citenamefont {Aggarwal},\ and\ \citenamefont {Dieckmann}()}]{topfer_point_1995}%
  \BibitemOpen
  \bibfield  {author} {\bibinfo {author} {\bibfnamefont {J.}~\bibnamefont {Töpfer}}, \bibinfo {author} {\bibfnamefont {S.}~\bibnamefont {Aggarwal}}, \ and\ \bibinfo {author} {\bibfnamefont {R.}~\bibnamefont {Dieckmann}},\ }\bibfield  {title} {\enquote {\bibinfo {title} {Point defects and cation tracer diffusion in ({CrxFe}1 − x)3 − δo4 spinels},}\ }\href {\doibase 10.1016/0167-2738(95)00190-H} {\ \textbf {\bibinfo {volume} {81}},\ \bibinfo {pages} {251--266}}\BibitemShut {NoStop}%
\bibitem [{\citenamefont {Howard}()}]{howard_random-walk_1966}%
  \BibitemOpen
  \bibfield  {author} {\bibinfo {author} {\bibfnamefont {R.~E.}\ \bibnamefont {Howard}},\ }\bibfield  {title} {\enquote {\bibinfo {title} {Random-walk method for calculating correlation factors: Tracer diffusion by divacancy and impurity-vacancy pairs in cubic crystals},}\ }\href {\doibase 10.1103/PhysRev.144.650} {\ \textbf {\bibinfo {volume} {144}},\ \bibinfo {pages} {650--661}},\ \bibinfo {note} {publisher: American Physical Society}\BibitemShut {NoStop}%
\bibitem [{\citenamefont {Klugkist}\ and\ \citenamefont {Herzig}()}]{klugkist_tracer_1995}%
  \BibitemOpen
  \bibfield  {author} {\bibinfo {author} {\bibfnamefont {P.}~\bibnamefont {Klugkist}}\ and\ \bibinfo {author} {\bibfnamefont {C.}~\bibnamefont {Herzig}},\ }\bibfield  {title} {\enquote {\bibinfo {title} {Tracer diffusion of titanium in α-iron},}\ }\href {\doibase 10.1002/pssa.2211480209} {\ \textbf {\bibinfo {volume} {148}},\ \bibinfo {pages} {413--421}},\ \bibinfo {note} {\_eprint: https://onlinelibrary.wiley.com/doi/pdf/10.1002/pssa.2211480209}\BibitemShut {NoStop}%
\bibitem [{\citenamefont {Mehrer}()}]{mehrer_diffusion_2007}%
  \BibitemOpen
  \bibfield  {author} {\bibinfo {author} {\bibfnamefont {H.}~\bibnamefont {Mehrer}},\ }\href@noop {} {\emph {\bibinfo {title} {Diffusion in Solids: Fundamentals, Methods, Materials, Diffusion-Controlled Processes}}}\ (\bibinfo  {publisher} {Springer Science \& Business Media})\ \bibinfo {note} {google-Books-{ID}: {IUZVffQLFKQC}}\BibitemShut {NoStop}%
\bibitem [{\citenamefont {Johansen}\ \emph {et~al.}()\citenamefont {Johansen}, \citenamefont {Vines}, \citenamefont {Bjørheim}, \citenamefont {Schifano},\ and\ \citenamefont {Svensson}}]{johansen_aluminum_2015}%
  \BibitemOpen
  \bibfield  {author} {\bibinfo {author} {\bibfnamefont {K.}~\bibnamefont {Johansen}}, \bibinfo {author} {\bibfnamefont {L.}~\bibnamefont {Vines}}, \bibinfo {author} {\bibfnamefont {T.}~\bibnamefont {Bjørheim}}, \bibinfo {author} {\bibfnamefont {R.}~\bibnamefont {Schifano}}, \ and\ \bibinfo {author} {\bibfnamefont {B.}~\bibnamefont {Svensson}},\ }\bibfield  {title} {\enquote {\bibinfo {title} {Aluminum migration and intrinsic defect interaction in single-crystal zinc oxide},}\ }\href {\doibase 10.1103/PhysRevApplied.3.024003} {\ \textbf {\bibinfo {volume} {3}},\ \bibinfo {pages} {024003}},\ \bibinfo {note} {publisher: American Physical Society}\BibitemShut {NoStop}%
\bibitem [{\citenamefont {Frodason}\ \emph {et~al.}({\natexlab{b}})\citenamefont {Frodason}, \citenamefont {Krzyzaniak}, \citenamefont {Vines}, \citenamefont {Varley}, \citenamefont {Van~de Walle},\ and\ \citenamefont {Johansen}}]{frodason_diffusion_2023}%
  \BibitemOpen
  \bibfield  {author} {\bibinfo {author} {\bibfnamefont {Y.~K.}\ \bibnamefont {Frodason}}, \bibinfo {author} {\bibfnamefont {P.~P.}\ \bibnamefont {Krzyzaniak}}, \bibinfo {author} {\bibfnamefont {L.}~\bibnamefont {Vines}}, \bibinfo {author} {\bibfnamefont {J.~B.}\ \bibnamefont {Varley}}, \bibinfo {author} {\bibfnamefont {C.~G.}\ \bibnamefont {Van~de Walle}}, \ and\ \bibinfo {author} {\bibfnamefont {K.~M.~H.}\ \bibnamefont {Johansen}},\ }\bibfield  {title} {\enquote {\bibinfo {title} {Diffusion of sn donors in β-ga2o3},}\ }\href {\doibase 10.1063/5.0142671} {\ \textbf {\bibinfo {volume} {11}},\ \bibinfo {pages} {041121} ({\natexlab{b}})}\BibitemShut {NoStop}%
\bibitem [{\citenamefont {Bracht}()}]{bracht_diffusion_2000}%
  \BibitemOpen
  \bibfield  {author} {\bibinfo {author} {\bibfnamefont {H.}~\bibnamefont {Bracht}},\ }\bibfield  {title} {\enquote {\bibinfo {title} {Diffusion mechanisms and intrinsic point-defect properties in silicon},}\ }\href {\doibase 10.1557/mrs2000.94} {\ \textbf {\bibinfo {volume} {25}},\ \bibinfo {pages} {22--27}}\BibitemShut {NoStop}%
\bibitem [{\citenamefont {Gösele}\ and\ \citenamefont {Tan}()}]{gosele_point_1991}%
  \BibitemOpen
  \bibfield  {author} {\bibinfo {author} {\bibfnamefont {U.~M.}\ \bibnamefont {Gösele}}\ and\ \bibinfo {author} {\bibfnamefont {T.~Y.}\ \bibnamefont {Tan}},\ }\bibfield  {title} {\enquote {\bibinfo {title} {Point defects and diffusion in semiconductors},}\ }\href {\doibase 10.1557/S0883769400055512} {\ \textbf {\bibinfo {volume} {16}},\ \bibinfo {pages} {42--46}}\BibitemShut {NoStop}%
\bibitem [{\citenamefont {Hobbs}\ and\ \citenamefont {Ord}()}]{hobbs_chapter_2015}%
  \BibitemOpen
  \bibfield  {author} {\bibinfo {author} {\bibfnamefont {B.}~\bibnamefont {Hobbs}}\ and\ \bibinfo {author} {\bibfnamefont {A.}~\bibnamefont {Ord}},\ }\bibfield  {title} {\enquote {\bibinfo {title} {Chapter 9 - visco-plastic flow},}\ }in\ \href {\doibase 10.1016/B978-0-12-407820-8.00009-6} {\emph {\bibinfo {booktitle} {Structural Geology}}},\ \bibinfo {editor} {edited by\ \bibinfo {editor} {\bibfnamefont {B.}~\bibnamefont {Hobbs}}\ and\ \bibinfo {editor} {\bibfnamefont {A.}~\bibnamefont {Ord}}}\ (\bibinfo  {publisher} {Elsevier})\ pp.\ \bibinfo {pages} {287--326}\BibitemShut {NoStop}%
\bibitem [{\citenamefont {Krasikov}\ and\ \citenamefont {Sankin}()}]{krasikov_beyond_2018}%
  \BibitemOpen
  \bibfield  {author} {\bibinfo {author} {\bibfnamefont {D.}~\bibnamefont {Krasikov}}\ and\ \bibinfo {author} {\bibfnamefont {I.}~\bibnamefont {Sankin}},\ }\bibfield  {title} {\enquote {\bibinfo {title} {Beyond thermodynamic defect models: A kinetic simulation of arsenic activation in {CdTe}},}\ }\href {\doibase 10.1103/PhysRevMaterials.2.103803} {\ \textbf {\bibinfo {volume} {2}},\ \bibinfo {pages} {103803}},\ \bibinfo {note} {publisher: American Physical Society}\BibitemShut {NoStop}%
\bibitem [{\citenamefont {Onsager}()}]{onsager_reciprocal_1931}%
  \BibitemOpen
  \bibfield  {author} {\bibinfo {author} {\bibfnamefont {L.}~\bibnamefont {Onsager}},\ }\bibfield  {title} {\enquote {\bibinfo {title} {Reciprocal relations in irreversible processes. i.}}\ }\href {\doibase 10.1103/PhysRev.37.405} {\ \textbf {\bibinfo {volume} {37}},\ \bibinfo {pages} {405--426}},\ \bibinfo {note} {publisher: American Physical Society}\BibitemShut {NoStop}%
\bibitem [{noa()}]{noauthor_matlab_2022}%
  \BibitemOpen
  \href {https://www.mathworks.com} {\enquote {\bibinfo {title} {{MATLAB} version: 9.13.0 (r2022b)},}\ }\BibitemShut {NoStop}%
\bibitem [{\citenamefont {Jacobs}()}]{jacobs_f_1974}%
  \BibitemOpen
  \bibfield  {author} {\bibinfo {author} {\bibfnamefont {K.}~\bibnamefont {Jacobs}},\ }\bibfield  {title} {\enquote {\bibinfo {title} {F. a. kröger. the chemistry of imperfect crystals. 2nd revised edition, volume 1: Preparation, purification, crystal growth and phase theory. north-holland publishing company - amsterdam/london 1973 american elsevier publishing company, inc. - new york 313 seiten, zahlreiche abbildungen und tabellen, kunstleder preis dfl. 70.00},}\ }\href {\doibase 10.1002/crat.19740090719} {\ \textbf {\bibinfo {volume} {9}},\ \bibinfo {pages} {K67--K68}},\ \bibinfo {note} {\_eprint: https://onlinelibrary.wiley.com/doi/pdf/10.1002/crat.19740090719}\BibitemShut {NoStop}%
\bibitem [{\citenamefont {Varley}()}]{varley_first-principles_2020}%
  \BibitemOpen
  \bibfield  {author} {\bibinfo {author} {\bibfnamefont {J.~B.}\ \bibnamefont {Varley}},\ }\bibfield  {title} {\enquote {\bibinfo {title} {First-principles calculations 2},}\ }in\ \href {\doibase 10.1007/978-3-030-37153-1_18} {\emph {\bibinfo {booktitle} {Gallium Oxide: Materials Properties, Crystal Growth, and Devices}}},\ \bibinfo {series and number} {Springer Series in Materials Science},\ \bibinfo {editor} {edited by\ \bibinfo {editor} {\bibfnamefont {M.}~\bibnamefont {Higashiwaki}}\ and\ \bibinfo {editor} {\bibfnamefont {S.}~\bibnamefont {Fujita}}}\ (\bibinfo  {publisher} {Springer International Publishing})\ pp.\ \bibinfo {pages} {329--348}\BibitemShut {NoStop}%
\bibitem [{\citenamefont {von Bardeleben}\ \emph {et~al.}()\citenamefont {von Bardeleben}, \citenamefont {He}, \citenamefont {Wu},\ and\ \citenamefont {Ding}}]{von_bardeleben_high_2023}%
  \BibitemOpen
  \bibfield  {author} {\bibinfo {author} {\bibfnamefont {H.~J.}\ \bibnamefont {von Bardeleben}}, \bibinfo {author} {\bibfnamefont {G.}~\bibnamefont {He}}, \bibinfo {author} {\bibfnamefont {Y.}~\bibnamefont {Wu}}, \ and\ \bibinfo {author} {\bibfnamefont {S.}~\bibnamefont {Ding}},\ }\bibfield  {title} {\enquote {\bibinfo {title} {High temperature annealing of n-type bulk β-ga2o3: Electrical compensation and defect analysis—the role of gallium vacancies},}\ }\href {\doibase 10.1063/5.0173581} {\ \textbf {\bibinfo {volume} {134}},\ \bibinfo {pages} {165702}}\BibitemShut {NoStop}%
\bibitem [{\citenamefont {Tuomisto}()}]{tuomisto_identification_2021}%
  \BibitemOpen
  \bibfield  {author} {\bibinfo {author} {\bibfnamefont {F.}~\bibnamefont {Tuomisto}},\ }\bibfield  {title} {\enquote {\bibinfo {title} {Identification of point defects in multielement compounds and alloys with positron annihilation spectroscopy: Challenges and opportunities},}\ }\href {\doibase 10.1002/pssr.202100177} {\ \textbf {\bibinfo {volume} {15}},\ \bibinfo {pages} {2100177}}\BibitemShut {NoStop}%
\bibitem [{\citenamefont {Ghadi}\ \emph {et~al.}()\citenamefont {Ghadi}, \citenamefont {{McGlone}}, \citenamefont {Jackson}, \citenamefont {Farzana}, \citenamefont {Feng}, \citenamefont {Bhuiyan}, \citenamefont {Zhao}, \citenamefont {Arehart},\ and\ \citenamefont {Ringel}}]{ghadi_full_2020}%
  \BibitemOpen
  \bibfield  {author} {\bibinfo {author} {\bibfnamefont {H.}~\bibnamefont {Ghadi}}, \bibinfo {author} {\bibfnamefont {J.~F.}\ \bibnamefont {{McGlone}}}, \bibinfo {author} {\bibfnamefont {C.~M.}\ \bibnamefont {Jackson}}, \bibinfo {author} {\bibfnamefont {E.}~\bibnamefont {Farzana}}, \bibinfo {author} {\bibfnamefont {Z.}~\bibnamefont {Feng}}, \bibinfo {author} {\bibfnamefont {A.~F. M. A.~U.}\ \bibnamefont {Bhuiyan}}, \bibinfo {author} {\bibfnamefont {H.}~\bibnamefont {Zhao}}, \bibinfo {author} {\bibfnamefont {A.~R.}\ \bibnamefont {Arehart}}, \ and\ \bibinfo {author} {\bibfnamefont {S.~A.}\ \bibnamefont {Ringel}},\ }\bibfield  {title} {\enquote {\bibinfo {title} {Full bandgap defect state characterization of β-ga2o3 grown by metal organic chemical vapor deposition},}\ }\href {\doibase 10.1063/1.5142313} {\ \textbf {\bibinfo {volume} {8}},\ \bibinfo {pages} {021111}},\ \bibinfo {note} {publisher: American Institute of Physics}\BibitemShut {NoStop}%
\bibitem [{\citenamefont {Johnson}\ \emph {et~al.}()\citenamefont {Johnson}, \citenamefont {Chen}, \citenamefont {Varley}, \citenamefont {Jackson}, \citenamefont {Farzana}, \citenamefont {Zhang}, \citenamefont {Arehart}, \citenamefont {Huang}, \citenamefont {Genc}, \citenamefont {Ringel}, \citenamefont {Van~de Walle}, \citenamefont {Muller},\ and\ \citenamefont {Hwang}}]{johnson_unusual_2019}%
  \BibitemOpen
  \bibfield  {author} {\bibinfo {author} {\bibfnamefont {J.~M.}\ \bibnamefont {Johnson}}, \bibinfo {author} {\bibfnamefont {Z.}~\bibnamefont {Chen}}, \bibinfo {author} {\bibfnamefont {J.~B.}\ \bibnamefont {Varley}}, \bibinfo {author} {\bibfnamefont {C.~M.}\ \bibnamefont {Jackson}}, \bibinfo {author} {\bibfnamefont {E.}~\bibnamefont {Farzana}}, \bibinfo {author} {\bibfnamefont {Z.}~\bibnamefont {Zhang}}, \bibinfo {author} {\bibfnamefont {A.~R.}\ \bibnamefont {Arehart}}, \bibinfo {author} {\bibfnamefont {H.-L.}\ \bibnamefont {Huang}}, \bibinfo {author} {\bibfnamefont {A.}~\bibnamefont {Genc}}, \bibinfo {author} {\bibfnamefont {S.~A.}\ \bibnamefont {Ringel}}, \bibinfo {author} {\bibfnamefont {C.~G.}\ \bibnamefont {Van~de Walle}}, \bibinfo {author} {\bibfnamefont {D.~A.}\ \bibnamefont {Muller}}, \ and\ \bibinfo {author} {\bibfnamefont {J.}~\bibnamefont {Hwang}},\ }\bibfield  {title} {\enquote {\bibinfo {title} {Unusual formation of point-defect complexes in the ultrawide-band-gap semiconductor
  \${\textbackslash}ensuremath\{{\textbackslash}beta\}{\textbackslash}text\{{\textbackslash}ensuremath\{-\}\}\{{\textbackslash}mathrm\{Ga\}\}\_\{2\}\{{\textbackslash}mathrm\{O\}\}\_\{3\}\$},}\ }\href {\doibase 10.1103/PhysRevX.9.041027} {\ \textbf {\bibinfo {volume} {9}},\ \bibinfo {pages} {041027}},\ \bibinfo {note} {publisher: American Physical Society}\BibitemShut {NoStop}%
\bibitem [{\citenamefont {Galazka}\ \emph {et~al.}()\citenamefont {Galazka}, \citenamefont {Uecker}, \citenamefont {Irmscher}, \citenamefont {Albrecht}, \citenamefont {Klimm}, \citenamefont {Pietsch}, \citenamefont {Brützam}, \citenamefont {Bertram}, \citenamefont {Ganschow},\ and\ \citenamefont {Fornari}}]{galazka_czochralski_2010}%
  \BibitemOpen
  \bibfield  {author} {\bibinfo {author} {\bibfnamefont {Z.}~\bibnamefont {Galazka}}, \bibinfo {author} {\bibfnamefont {R.}~\bibnamefont {Uecker}}, \bibinfo {author} {\bibfnamefont {K.}~\bibnamefont {Irmscher}}, \bibinfo {author} {\bibfnamefont {M.}~\bibnamefont {Albrecht}}, \bibinfo {author} {\bibfnamefont {D.}~\bibnamefont {Klimm}}, \bibinfo {author} {\bibfnamefont {M.}~\bibnamefont {Pietsch}}, \bibinfo {author} {\bibfnamefont {M.}~\bibnamefont {Brützam}}, \bibinfo {author} {\bibfnamefont {R.}~\bibnamefont {Bertram}}, \bibinfo {author} {\bibfnamefont {S.}~\bibnamefont {Ganschow}}, \ and\ \bibinfo {author} {\bibfnamefont {R.}~\bibnamefont {Fornari}},\ }\bibfield  {title} {\enquote {\bibinfo {title} {Czochralski growth and characterization of β-ga2o3 single crystals},}\ }\href {\doibase 10.1002/crat.201000341} {\ \textbf {\bibinfo {volume} {45}},\ \bibinfo {pages} {1229--1236}},\ \bibinfo {note} {\_eprint: https://onlinelibrary.wiley.com/doi/pdf/10.1002/crat.201000341}\BibitemShut {NoStop}%
\bibitem [{\citenamefont {Kuramata}\ \emph {et~al.}()\citenamefont {Kuramata}, \citenamefont {Koshi}, \citenamefont {Watanabe}, \citenamefont {Yamaoka}, \citenamefont {Masui},\ and\ \citenamefont {Yamakoshi}}]{kuramata_high-quality_2016}%
  \BibitemOpen
  \bibfield  {author} {\bibinfo {author} {\bibfnamefont {A.}~\bibnamefont {Kuramata}}, \bibinfo {author} {\bibfnamefont {K.}~\bibnamefont {Koshi}}, \bibinfo {author} {\bibfnamefont {S.}~\bibnamefont {Watanabe}}, \bibinfo {author} {\bibfnamefont {Y.}~\bibnamefont {Yamaoka}}, \bibinfo {author} {\bibfnamefont {T.}~\bibnamefont {Masui}}, \ and\ \bibinfo {author} {\bibfnamefont {S.}~\bibnamefont {Yamakoshi}},\ }\bibfield  {title} {\enquote {\bibinfo {title} {High-quality β-ga2o3 single crystals grown by edge-defined film-fed growth},}\ }\href {\doibase 10.7567/JJAP.55.1202A2} {\ \textbf {\bibinfo {volume} {55}},\ \bibinfo {pages} {1202A2}},\ \bibinfo {note} {publisher: {IOP} Publishing}\BibitemShut {NoStop}%
\bibitem [{\citenamefont {Zhang}\ \emph {et~al.}()\citenamefont {Zhang}, \citenamefont {Farzana}, \citenamefont {Arehart},\ and\ \citenamefont {Ringel}}]{zhang_deep_2016}%
  \BibitemOpen
  \bibfield  {author} {\bibinfo {author} {\bibfnamefont {Z.}~\bibnamefont {Zhang}}, \bibinfo {author} {\bibfnamefont {E.}~\bibnamefont {Farzana}}, \bibinfo {author} {\bibfnamefont {A.~R.}\ \bibnamefont {Arehart}}, \ and\ \bibinfo {author} {\bibfnamefont {S.~A.}\ \bibnamefont {Ringel}},\ }\bibfield  {title} {\enquote {\bibinfo {title} {Deep level defects throughout the bandgap of (010) β-ga2o3 detected by optically and thermally stimulated defect spectroscopy},}\ }\href {\doibase 10.1063/1.4941429} {\ \textbf {\bibinfo {volume} {108}},\ \bibinfo {pages} {052105}},\ \bibinfo {note} {publisher: American Institute of Physics}\BibitemShut {NoStop}%
\bibitem [{\citenamefont {Ingebrigtsen}\ \emph {et~al.}({\natexlab{b}})\citenamefont {Ingebrigtsen}, \citenamefont {Varley}, \citenamefont {Kuznetsov}, \citenamefont {Svensson}, \citenamefont {Alfieri}, \citenamefont {Mihaila}, \citenamefont {Badstübner},\ and\ \citenamefont {Vines}}]{ingebrigtsen_iron_2018}%
  \BibitemOpen
  \bibfield  {author} {\bibinfo {author} {\bibfnamefont {M.~E.}\ \bibnamefont {Ingebrigtsen}}, \bibinfo {author} {\bibfnamefont {J.~B.}\ \bibnamefont {Varley}}, \bibinfo {author} {\bibfnamefont {A.~Y.}\ \bibnamefont {Kuznetsov}}, \bibinfo {author} {\bibfnamefont {B.~G.}\ \bibnamefont {Svensson}}, \bibinfo {author} {\bibfnamefont {G.}~\bibnamefont {Alfieri}}, \bibinfo {author} {\bibfnamefont {A.}~\bibnamefont {Mihaila}}, \bibinfo {author} {\bibfnamefont {U.}~\bibnamefont {Badstübner}}, \ and\ \bibinfo {author} {\bibfnamefont {L.}~\bibnamefont {Vines}},\ }\bibfield  {title} {\enquote {\bibinfo {title} {Iron and intrinsic deep level states in ga2o3},}\ }\href {\doibase 10.1063/1.5020134} {\ \textbf {\bibinfo {volume} {112}},\ \bibinfo {pages} {042104} ({\natexlab{b}})}\BibitemShut {NoStop}%
\bibitem [{\citenamefont {Crank}()}]{crank_mathematics_1956}%
  \BibitemOpen
  \bibfield  {author} {\bibinfo {author} {\bibfnamefont {J.}~\bibnamefont {Crank}},\ }\enquote {\bibinfo {title} {The mathematics of diffusion},}\ \BibitemShut {NoStop}%
\bibitem [{\citenamefont {{McCloy}}\ \emph {et~al.}()\citenamefont {{McCloy}}, \citenamefont {Jesenovec}, \citenamefont {Dutton}, \citenamefont {Pansegrau}, \citenamefont {Remple}, \citenamefont {Weber}, \citenamefont {Swain}, \citenamefont {{McCluskey}},\ and\ \citenamefont {Scarpulla}}]{mccloy_growth_2022}%
  \BibitemOpen
\bibfield  {title} {  }\bibfield  {author} {\bibinfo {author} {\bibfnamefont {J.~S.}\ \bibnamefont {{McCloy}}}, \bibinfo {author} {\bibfnamefont {J.}~\bibnamefont {Jesenovec}}, \bibinfo {author} {\bibfnamefont {B.}~\bibnamefont {Dutton}}, \bibinfo {author} {\bibfnamefont {C.}~\bibnamefont {Pansegrau}}, \bibinfo {author} {\bibfnamefont {C.}~\bibnamefont {Remple}}, \bibinfo {author} {\bibfnamefont {M.~H.}\ \bibnamefont {Weber}}, \bibinfo {author} {\bibfnamefont {S.}~\bibnamefont {Swain}}, \bibinfo {author} {\bibfnamefont {M.}~\bibnamefont {{McCluskey}}}, \ and\ \bibinfo {author} {\bibfnamefont {M.}~\bibnamefont {Scarpulla}},\ }\bibfield  {title} {\enquote {\bibinfo {title} {Growth and defect characterization of doped and undoped β-ga2o3 crystals},}\ }in\ \href {\doibase 10.1117/12.2611852} {\emph {\bibinfo {booktitle} {Oxide-based Materials and Devices {XIII}}}},\ \bibinfo {editor} {edited by\ \bibinfo {editor} {\bibfnamefont {F.~H.}\ \bibnamefont {Teherani}}\ and\ \bibinfo {editor} {\bibfnamefont {D.~J.}\
  \bibnamefont {Rogers}}}\ (\bibinfo  {publisher} {{SPIE}})\ p.~\bibinfo {pages} {10}\BibitemShut {NoStop}%
\bibitem [{\citenamefont {Uhlendorf}\ and\ \citenamefont {Schmidt}()}]{uhlendorf_oxygen_2023}%
  \BibitemOpen
  \bibfield  {author} {\bibinfo {author} {\bibfnamefont {J.}~\bibnamefont {Uhlendorf}}\ and\ \bibinfo {author} {\bibfnamefont {H.}~\bibnamefont {Schmidt}},\ }\bibfield  {title} {\enquote {\bibinfo {title} {Oxygen and aluminum tracer diffusion in \$({\textbackslash}ensuremath\{-\}201)\$ oriented \${\textbackslash}ensuremath\{{\textbackslash}beta\}{\textbackslash}text\{{\textbackslash}ensuremath\{-\}\}{\textbackslash}mathrm\{G\}\{{\textbackslash}mathrm\{a\}\}\_\{2\}\{{\textbackslash}mathrm\{O\}\}\_\{3\}\$ single crystals},}\ }\href {\doibase 10.1103/PhysRevMaterials.7.093402} {\ \textbf {\bibinfo {volume} {7}},\ \bibinfo {pages} {093402}},\ \bibinfo {note} {publisher: American Physical Society}\BibitemShut {NoStop}%
\bibitem [{\citenamefont {Azarov}\ \emph {et~al.}()\citenamefont {Azarov}, \citenamefont {Venkatachalapathy}, \citenamefont {Vines}, \citenamefont {Monakhov}, \citenamefont {Lee},\ and\ \citenamefont {Kuznetsov}}]{azarov_activation_2021}%
  \BibitemOpen
  \bibfield  {author} {\bibinfo {author} {\bibfnamefont {A.}~\bibnamefont {Azarov}}, \bibinfo {author} {\bibfnamefont {V.}~\bibnamefont {Venkatachalapathy}}, \bibinfo {author} {\bibfnamefont {L.}~\bibnamefont {Vines}}, \bibinfo {author} {\bibfnamefont {E.}~\bibnamefont {Monakhov}}, \bibinfo {author} {\bibfnamefont {I.-H.}\ \bibnamefont {Lee}}, \ and\ \bibinfo {author} {\bibfnamefont {A.}~\bibnamefont {Kuznetsov}},\ }\bibfield  {title} {\enquote {\bibinfo {title} {Activation energy of silicon diffusion in gallium oxide: Roles of the mediating defects charge states and phase modification},}\ }\href {\doibase 10.1063/5.0070045} {\ \textbf {\bibinfo {volume} {119}},\ \bibinfo {pages} {182103}},\ \bibinfo {note} {publisher: American Institute of Physics}\BibitemShut {NoStop}%
\end{thebibliography}%

\end{document}